# Testing and Verification of Quantum Compilers through Assurance Contracts and Evidence

Running title: Quantum compiler testing and verification

Furqan Nasir[a],[b], Muhammad Arif Shah[a], Iftikhar Alam[a]

[a] City University of Science and Information Technology (CUSIT), Peshawar, Pakistan

[b] National University of Computer and Emerging Sciences (FAST-NUCES), Islamabad, Pakistan

Correspondence: Furqan Nasir, City University of Science and Information Technology, Dalazak Road, Peshawar, Pakistan. Email: furqannr@gmail.com

ORCID: Furqan Nasir https://orcid.org/0000-0002-5259-3448; Muhammad Arif Shah https://orcid.org/0000-0003-0090-3333; Iftikhar Alam https://orcid.org/0000-0002-8087-5485

## Abstract

Quantum compiler assurance requires a semantic relation that matches the intended use and observations that can expose violations at the delivered interface. This critical integrative survey compares verified transformations, equivalence checking, differential and metamorphic testing, property-based testing, and evidence for assurance across revisions. A source catalogue and focused extraction matrix distinguish formal guarantees, reported empirical findings, analytical deductions, and proposed practice. A recorded coverage update and source-level selection decisions make the review auditable within its declared scope. Four documented failure cases connect conditional applicability, parameter association, termination, and phase conventions to method selection. A worked example involving layout metadata, parameter bindings, and ancillary qubits demonstrates why validating a circuit under a reconstructed interpretation can leave the delivered interface unchecked. An illustrative revision scenario then separates fault detection from change attribution. The resulting guidance specifies applicability, complementary observations, and treatment of inconclusive outcomes. Published bug counts, mutation results, and processing benchmarks support different claims and do not establish a common effectiveness ranking. The proposed regression architecture remains unevaluated as a complete system; its practical value requires controlled comparisons on independently adjudicated compiler-change events.

## 1 Introduction

Quantum compilers must preserve program meaning while producing executable artifacts whose layout, parameter bindings, and result records support correct use. These obligations interact with target constraints and optimization objectives. A semantically valid circuit can be delivered with an incorrect interpretation record, while deterioration in one resource metric can be an intentional trade-off. An assurance claim must therefore identify both the requirement and the evidence used to assess it.

Compilation is a sequence of transformations rather than one operation. Gate decomposition, placement, routing, optimization, and scheduling each alter the circuit, and some of those changes also alter how the circuit must be interpreted. Existing reviews already describe these stages and the points at which their abstractions break down [1, 2]. The assurance problem is harder because the semantics are quantum. A phase that is irrelevant for an isolated operation can become observable as soon as the same operation is used under coherent control. Likewise, a partially initialized input or a particular measurement choice may require its own equivalence relation, while a parameterized optimization may need to be correct over an entire domain rather than at one setting. Dynamic circuits add classical records alongside the quantum state [3, 4, 5]. These are not peripheral details. Whether a pass, failure, or inconclusive verdict actually supports the attached quality claim depends on treating these distinctions correctly.

The main assurance approaches do not provide the same form of evidence. A verified optimizer or compiler pass proves preservation within a formal model [6, 7]. An equivalence checker instead validates a particular pair of circuits through an algebraic or compact representation [8, 9]. A proof and a pairwise check therefore support different claims. Differential and metamorphic testing take another route by executing related program variants and looking for discrepancies or failures [10, 11, 12, 13]. Property-based testing evaluates a supplied property over generated inputs [14]. Benchmarks answer a different question again: they measure compilation cost or output quality, not correctness [15]. Comparing these approaches is useful only if the comparison keeps their real differences visible - how inputs are quantified, which program features are supported, what behaviour is observed, and on which workloads the evidence was obtained.

Compiler quality and test quality are separate assessment targets. Fault detection, reliability, and resource efficiency are attributes of a test suite as well as concerns of the system under test [16]. An unstable oracle can reject a correct compiler, while an expensive checker can leave substantial input domains unexamined. The classical oracle problem provides a foundation for this distinction [17]; quantum compiler assurance additionally requires an explicit semantic relation at each compilation boundary.

Prior syntheses cover quantum testing and analysis, debugging, empirical methodology, and tool support [18, 19, 20, 21, 22]. Measurement-based and statevector-based validation have also been compared directly [23]. The contribution here is a compiler-specific connection between the preservation obligation, the evidence obtainable at the delivered interface, and the interpretation of that evidence across revisions. Combining testing and verification, or observing limitations of measurement, is not claimed as a new idea.

The survey contributes three results. First, it compares semantic relations and interface obligations under a common contract description. Second, it separates guarantees, empirical findings, and analytical implications in a source-linked evidence matrix. Third, it applies a conditional selection procedure to documented failure mechanisms and a worked interface example. The procedure identifies what a method can establish, what additional observation is needed, and when the result must remain inconclusive. Its implications for regression evaluation are proposed practice, not a measured effectiveness claim.

The paper is organized as follows. Section 2 sets out how the review was done. Section 3 fixes the requirements and the semantic boundaries. Sections 4 and 5 put verification evidence and testing evidence side by side. Section 6 then walks through the documented cases and the findings that run across them, and Sections 7 and 8 turn those findings into concrete guidance, one on selecting methods and the other on what a regression evaluation has to report. Section 9 presents research priorities and threats to validity; Section 10 concludes.

## 2 Review design and evidence assessment

### 2.1 Scope and questions

This review critically integrates evidence on quality assurance for gate-based quantum compilers. Its unit of analysis is an assurance mechanism together with the claim that mechanism can support, rather than the number of papers discussing it. The scope includes transformations of unitary and parameterized circuits, initialization and measurement, target mapping, output interpretation, and validation across compiler revisions. Program-testing work is included when its oracle or evaluation mechanism can be transferred to compilation. Classical testing studies are also relevant where they inform effort allocation or the interpretation of regression evidence. Pulse synthesis, error-correcting-code implementation, hardware characterization, cryptographic security, and application-specific optimization remain outside the primary scope and are considered only when they help define an assurance boundary.

Five research questions organize the synthesis. RQ1 concerns operational compiler requirements; RQ2 concerns semantic relations, input domains, and guarantees; RQ3 concerns the meaning of published evaluation evidence; RQ4 concerns conditional complementarity among methods; and RQ5 concerns the evidence required to extend assurance across revisions. Sections 3, 4, 5, 6–7, and 8–9 address these questions respectively.

## 2.2 Identification and scope control

The review covers assurance of gate-model quantum compilation: verified transformations, translation validation, executable testing, interface interpretation, and evidence across revisions. Program-level methods are included when their semantic relation or observation mechanism transfers to a compilation boundary. Classical regression studies provide methodological context and are not treated as quantum-compiler evaluations. Optimization-only work without an assurance mechanism, hardware-only verification, post-quantum cryptography, and application studies without a relevant assurance boundary are outside scope.

The initial study set originated in preliminary work for an SLR of testing and regression testing of quantum compilers. Its protocol planned English-language studies from 2015–2026 across IEEE Xplore, ACM Digital Library, ScienceDirect, SpringerLink, Scopus, Web of Science, arXiv, and Google Scholar. Search strings combined compiler/transpiler terminology with testing, faults, oracles, equivalence, regression, and selection terms. A preliminary report describes a pilot on 9 July 2026 using IEEE Xplore, SpringerLink, and arXiv, with the other five sources inaccessible automatically. The report and populated extraction workbook identify seven seed studies: QDiff, MorphQ, QuteFuzz, QEMI, Giallar, CertiQ, and Bugs4Q. Supporting Information 1 preserves the planned queries and maps these seeds to the current publication records, including later versions.

The planned full SLR was not completed. A later July 2026 interim workbook extended the pilot to 42 provisionally included study families (S1–S6 and S8–S43), while S7 was excluded from that narrower review. Its screening, extraction and quality decisions include AI-assisted preliminary work that still required researcher confirmation; its reported agreement is between AI-assisted streams, not independent human reviewers. We used the interim families as a discovery cross-check and applied this survey's eligibility rules to individual publications. The survey also broadened scope to transferable program verification, circuit assurance, and classical regression methodology, so the historical SLR decisions are not imported as final survey decisions. The original pilot report provides aggregate counts, but its database exports and record-level screening decisions are unavailable; its earlier workbook's search-run fields are unfilled. Neither those pilot totals nor the later provisional workbook constitute a verified selection flow for this survey. Registration, independent duplicate human screening, and a completed systematic search are not claimed.

Identification then continued through topic and exact-title searches with selective reference following recorded through 13 September 2026, followed by two dated coverage updates. Record-level historical retrieval and exclusion counts remain unavailable. On 19 September 2026, six prospectively recorded query families covered compiler faults, verified compilation, equivalence, interfaces, regression, and neighboring surveys through the OpenAI web search service. Supporting Information 1 preserves the protocol, exact queries, returned records, and decisions. The service does not expose its underlying engine or complete ranked hit lists; archived return counts are therefore not bibliographic population counts.

The July interim workbook and its source notes are archived with hashes in Supporting Information 1. A 43-row lineage table maps its S1–S43 identifiers to current survey records or records a separate survey-screening disposition. This checks coverage without treating the workbook as a completed SLR or counting the same publication twice.

A broader archive of the July SLR subfolder retains search exports, screening packets, extraction and assessment records, together with a file inventory. It includes H1/H2-labelled title/abstract packets for 1,919 records, but the 600 disagreement/unclear adjudications explicitly identify an AI-assisted adjudicator pending lead-researcher review; the 537-record full-text eligibility register is blank in its decision field. Consequently, these historical files demonstrate substantial preparatory work but do not establish a completed human-adjudicated selection or full-text inclusion flow for this survey. The archive notes document the conflicting status labels and distinguish the historical narrow SLR boundary from this survey's broader eligibility rule.

Alternative terminology was checked explicitly: compiler and transpiler; equivalence checking and translation validation; control flow and hybrid or dynamic circuits; parameterized angles and circuit-size families; regression and transpilation drift. Three domain-restricted follow-up queries, primary-source follow-ups, and selected reference checks supplemented the six queries. In particular, the parameterized-equivalence literature led to cutoff theorems and rewrite-rule synthesis [24, 25], while the update identified shallow-circuit checking and a baseline-relative drift workflow [26, 27]. Quantum-inspired optimization of classical test suites was distinguished from testing quantum compilers. Each substantive new candidate has a recorded retention, contextual-use, or exclusion reason.

An additional coverage check on 20 September 2026 queried the public Crossref works API. Three successful queries returned the first 100 relevance-ranked records each (300 distinct DOIs); a fourth query was rate-limited. Five DBLP API queries produced no usable records. Crossref is a metadata service, and its broad relevance matching returned substantial off-topic material. These results support a bounded coverage check, not exhaustive database retrieval. All 300 returned records have title-level dispositions, and the recorded failures and query caps are retained in Supporting Information 1. Selected references in Paltenghi and Pradel [18] and Li et al. [21] were also followed, adding non-unitary validation, black-box relation testing, combinatorial generation, and integration-testing context. Exact-title follow-ups checked publication families; no complete forward-citation search is claimed.

A targeted follow-up on 21 September 2026 checked verified compiler pipelines, metamorphic replication, and mutation evidence against primary texts. It added ReVerC, ReQWIRE, VQO, and QBlue, and expanded extraction of CertiQ, MorphQ++, and the 2025 quantum-mutant study. Exact follow-up queries and source-level decisions appear in Supporting Information 1 and 2. This improves coverage of proof-to-test boundaries and approximation contracts without converting the bounded review into an exhaustive search.

A further exact-title check on 22 September 2026 used the unfinished SLR lineage to identify five compiler-assurance methods that the earlier catalogue had not represented separately: RelOpt, original AutoQ, Clifford-U equivalence, QuBEC, and improved decision-diagram checking [28, 29, 30, 31, 32]. Primary publisher or preprint pages were checked for publication identity and the limited scope claims used here. All five are marked scope-only in Supporting Information 2; their inclusion does not imply full technical extraction, final SLR eligibility, or measured compiler-regression effectiveness. The literature-publication cutoff remains 20 September 2026.

Publication families are linked by title, authors, DOI, and version. The 2021 VOQC study and its 2023 extension remain one lineage [6, 33]. QEMI uses the FASE 2026 publication [12]. QuCheck mechanism descriptions remain tied to the inspected earlier manuscript, with the 2026 online record identified separately [14]. Preprints and engineering reports retain explicit status. The catalogue contains 102 records, including related versions and contextual sources; it is not a census of independent methods.

### 2.3 Extraction and comparison

The evidence matrix contains 76 primary-method and evaluation records: 39 detailed scoped extractions and 37 scope-only records. One eligibility rule applies to both testing and verification: a primary mechanism or evaluation is included when it informs a semantic relation, observation, applicability condition, or evidence boundary used in the synthesis. Direct compiler-regression detection is a classification dimension, not an inclusion prerequisite. This includes transferable program methods such as projective-measurement testing and Qafny as well as compiler-facing methods. Reviews, classical transfer sources, engineering records, and a linked journal extension remain contextual. Scope-only fields are explicitly unassessed; their presence does not support performance comparisons or capability-absence claims.

Extraction records the assessed object, requirement, oracle inputs, relation and quantification, supported features, trust boundary, costs, evaluation setting, source locator, inspected version, and access depth. A separate evidence code distinguishes formal guarantees, reported empirical findings, analytical deductions, and proposed practice. A relevance code distinguishes compiler faults, compiler-change monitoring, quantum-program or circuit assurance, and transferred classical methodology. Direct detection of a compiler regression additionally requires an identified satisfying baseline and a violating candidate under a common contract; defect discovery alone does not qualify.

The finding ledger in Supporting Information 2 connects the conclusions in Section 6.5 to primary passages and identifies the step from source finding to review inference. Numerical results remain tied to their original workloads and versions. Abstract-only records support only their stated scope, not uninspected experimental details. The pilot's aggregate quality scores are retained as historical records; the present review instead applies the mechanism-based eligibility and source-specific evidence assessment above. Screening and extraction were not independently duplicated, and no cited experiments were rerun. These restrictions also apply to the authors' own recent preprint [34], which is identified as related work and is not independent corroboration of this survey.

### 2.4 Relation to earlier reviews

Earlier mapping and formal-method reviews provide additional context [35, 36]. Table 1 compares the organizing units of neighboring reviews with the compiler-level decision examined here. The comparison identifies overlap explicitly and does not infer that a topic is absent merely because it is not a review’s main organizing axis.

**Table 1** Relationship to earlier reviews

| Review | Established organizing perspective | Shared subject | Additional question addressed here |
|---|---|---|---|
| Cardama et al. [1, 2] | Compilation stages and intermediate representations | Transformation boundaries | What evidence validates a boundary under its declared interpretation? |
| Paltenghi and Pradel [18], Secs. 4–7 | Input generation, assertions, proof methods, abstractions, optimization, and datasets | Broad method and property coverage | Attach each mechanism to a compiler obligation and delivered interface |
| Ramalho et al. [19] | Testing and debugging research directions | Testing mechanisms and open problems | Which proposed regression practices have direct quantum evidence? |
| Li et al. [21], Sec. 3.1 and RQ1–RQ10 | Programs, mutants, scale, inputs, oracles, metrics, baselines, repetitions, backends, artifacts | Evaluation design and oracle limitations | Separate evidence of a defect from evidence of deterioration at a compiler-change boundary |
| Rosa and Santiago [20] | Assertions and debugging | Runtime observations and diagnosis | How are source-to-output transformations and their interfaces assured? |
| Ye et al. [23] | Measurement and statevector validation | Information available to an oracle | How do input quantification, mapping, and composition further restrict a verdict? |
| Wang et al. [22] and Andrews et al. [37] | Tool capabilities and functional validation | Executable assurance mechanisms | Which combinations add evidence under compatible requirements? |

| Review | Established organizing perspective | Shared subject | Additional question addressed here |
|---|---|---|---|
| Gil-Pereira et al. [38] | Research and practitioner perspectives | Testing practice | How can provenance support reproducible change attribution? |

RQ: research question.

The distinction is analytical rather than terminological. Paltenghi and Pradel organize techniques around testing and analysis tasks, including proof properties and datasets [18, Secs. 4–7]. Li et al. organize empirical practice through ten questions covering subjects, inputs, oracles, metrics, and reproducibility [21, Sec. 3.1; inspected arXiv v2]. This survey instead carries one compiler obligation through the relation, applicability check, observable result, and revision-attribution decision. Sections 7.4–7.5 demonstrate that procedure on layout, conditional-gate, and phase-sensitive examples. A method may be strong on one of these obligations and inapplicable to another.

### 2.5 Assessment limits

The review supports conditional relationships among the included mechanisms, not field-wide prevalence or a validated effectiveness ranking. Search caps, historical counts without underlying record-level logs, access restrictions, and the absence of independent duplicate extraction limit coverage and interpretation. Section 9.3 discusses these threats; source-specific boundaries accompany the evidence records.

## 3 Operational requirements and quantum semantic boundaries

### 3.1 A compilation result is more than a circuit

We model compilation as a function of a circuit C, backend description b, configuration k, and environment E. Configuration includes seeds and pass settings; the environment includes dependency versions and execution resources. Compilation returns a circuit C′, interpretation metadata M, termination or validity status S, resource measurements R, and trace T:

$$Compile(C, b, k, E) \rightarrow O = (C', M, S, R, T) \quad (1)$$

This formulation separates the compiled artifact from its interpretation and from the context in which it is executed. If an implementation omits traces or resource measurements, a downstream checker simply has less information available. An oracle o receives a projection $\pi_o(O)$, together with a reference record and specification when those are applicable, and returns one of three outcomes: pass, fail, or explicitly inconclusive. A timeout inside the oracle cannot be treated as semantic equivalence. The same separation is needed for other outcomes: an unsupported operation is different from a malformed circuit, and both are different from a demonstrated behavioural mismatch.

A requirement becomes operational only after five items are explicit: who owns it, the domain to which it applies, the rule for accepting an output, the observation on which that rule depends, and the treatment of uncertainty. Semantic obligations come from the compiler specification, while executable instructions and connectivity come from the target description. Developers or users must set resource budgets and optimization objectives. None of these requirements can be reconstructed reliably from a discrepancy observed after the fact. The same applies when the specification changes: compatibility with the new specification has to be decided before a difference between versions can be labelled a regression.

Table 2 separates correctness requirements from objectives that depend on context. Its categories classify obligations. They are not intended to imply that root causes are mutually exclusive. Metadata belongs to semantic correctness whenever it determines the logical interpretation of a compiled result. Reproducibility needs a further distinction. It may be a product requirement, such as deterministic output under a fixed seed, or an evaluation requirement, such as retaining enough context to interpret a comparison. Those two meanings should be recorded separately.

**Table 2** Operational compiler requirements and decision rules

| Requirement and authority | Applicable observation | Acceptance rule | Boundary and interaction |
|---|---|---|---|
| Semantic preservation from language and pass specifications | Interpreted source and output processes | Declared relation for the permitted inputs and contexts | Phase, initialization, measurement, and approximation determine the relation |
| Validity from the input language and target specification | Input eligibility, termination, instructions and connectivity | Accepted inputs produce well-formed supported outputs or documented diagnostics | Unsupported input and checker timeout are separate outcomes |
| Interface correctness from the compiler API | Layout, parameter binding and output associations | Consistency with independently reconstructed interpretation | A metadata violation can also be a semantic violation |
| Output quality from a declared optimization objective | Depth, two-qubit gates or other justified metrics on a matched target | Stated bound, permitted trade-off, or comparison criterion | A worse individual metric does not necessarily violate the objective |
| Resource use from an operational budget | Time, memory and resource configuration | Budget or distributional performance criterion fixed before evaluation | Infrastructure variation and workload changes need control |
| Reproducibility from product or evaluation requirements | Seeds, versions, target snapshot and normalized outputs | Declared deterministic or statistical stability condition | Different valid stochastic outputs need not be faults |

API: application programming interface.

One possible decision policy treats semantic and validity requirements, including any declared approximation allowance, as feasibility conditions. Optimization outcomes are then compared among outputs satisfying those conditions, using an explicit trade-off policy or Pareto comparison. Checker resource limits remain distinct from compiler performance: an incomplete semantic check establishes an unresolved obligation, not a semantic failure. This ordering is proposed guidance derived from the synthesis, not a universal compiler standard.

### 3.2 Quantum processes and observable distributions

A closed n-qubit computation is represented by a unitary U on a Hilbert space of dimension $2^n$. For input density operator ρ, the output is $U\rho U^\dagger$. A measurement with positive operators $E_\gamma$ satisfying $\Sigma_\gamma E_\gamma = I$ produces probabilities $p(y|\rho) = \mathrm{Tr}(E_\gamma U\rho U^\dagger)$. Both the input states and measurement operators restrict an executable oracle. Exact probability calculation removes sampling uncertainty for those observations, but does not enlarge the observed input or measurement domain [39, 40].

Ancillas alter the comparison domain. If compilation initializes auxiliary register A to $|0\rangle$ and later discards it, the logical output is obtained by tracing A out of $V(\rho \otimes |0\rangle\langle 0|)V^\dagger$ after accounting for layout. Correctness then requires equality of reduced processes on the permitted inputs. Equal-width full-matrix comparison is inappropriate, while requiring the ancilla to be returned clean imposes an additional reuse obligation. Selected measurement probabilities define a weaker relation than retained-state equality and cannot replace a reduced-channel contract [3].

Measurement and classical control require preservation of observable records as well as conditional states. Let each branch $\Phi_\gamma$ be a completely positive, trace-nonincreasing map, with $\Sigma_\gamma\Phi_\gamma$ trace preserving. Retaining the classical result gives $\Sigma_\gamma |y\rangle\langle y| \otimes \Phi_\gamma(\rho)$; discarding it gives $\Sigma_\gamma\Phi_\gamma(\rho)$. Equality of the retained classical-quantum output for every permitted ρ is stronger than equality after discarding the record. Implementations can therefore agree on the averaged state while associating a conditional state with the wrong result label. A hybrid comparison must preserve the record whenever the interface exposes it [41].

Approximation adds a quantitative condition to the relation already being checked. Any allowed synthesis error has to be stated in a particular metric and over a declared domain, with numerical error tracked separately. Agreement on finitely sampled probabilities, satisfaction of a channel-fidelity threshold, and a bound on operator distance are different premises and support different conclusions. Noisy execution adds another boundary: differences introduced by compilation must be distinguished from differences introduced by the physical process that realizes the compiled circuit [42, 43]. Table 3 links these quantum features to the comparisons they require, while Fig. 1 illustrates how measurement choice and reuse context determine which differences are observable.

**Table 3** Quantum compilation features and the appropriate comparison boundary

| Compilation feature | Appropriate relation or observation | Comparison error to avoid |
|---|---|---|
| Isolated unitary transformation | Operator equivalence under a declared global-phase convention | Applying the same phase convention to a controlled subroutine |
| Placement and routing | Logical input and output order reconstructed from mappings | Comparing physical gate lists without their interpretation |
| Initialized and discarded ancillas | Reduced output process when the quantum subsystem remains observable | Requiring equal-width full-unitary equality |
| Partial observational equivalence [3] | Selected measurement probabilities for all allowed inputs | Treating matching probabilities as equality of retained quantum states |
| Parameterized transformations [4] | Universal identity or explicitly binding-specific agreement | Treating finitely tested values as an unrestricted identity |
| Terminal measurements [40, 39] | Outcome probabilities for the chosen inputs and observables | Treating one measurement basis as complete state information |
| Measurement and feed-forward [41] | Classical outcomes and conditional quantum behavior | Comparing only the averaged quantum state |
| Approximate or noisy transformations [42, 43] | Declared process metric, tolerance, and noise assumptions | Interchanging exact equality and approximate agreement |

## 3.3 Semantic equivalence requires a declared domain

For equal-width closed unitary circuits, the standard relation is equality up to global phase, $V = \exp(i\varphi)U$ for some real $\varphi$. That relation is meaningful only if the checker interprets qubit order and layout consistently. Let $P_i$ map logical inputs into the compiled ordering and $P_o$ map logical outputs into the same ordering. The normalized compiled operator is then $P_o^{-1}VP_i$, which already assumes equal-width permutation maps. If compilation initializes ancillas or discards outputs, simple permutations are no longer enough. Embeddings and reduced channels are required instead. Applying a full-matrix comparison without that adaptation can therefore test a condition that is unnecessarily strong or simply wrong.

Phase equivalence depends on composition. The single-qubit operators I and −I define the same isolated channel, but controlled-(−I) applies Z to the control whereas controlled-I is identity. An interface permitting coherent reuse can therefore require phase information that an isolated-operation interface may discard. The checker’s normalization must follow the declared composition contract.

A fixed-input state comparison does not establish universal operator equivalence. Measurement can further restrict the observation: I and Z applied to $|+\rangle$ yield $|+\rangle$ and $|-\rangle$, which have identical computational-basis distributions but distinct X-basis outcomes. Additional shots in the original basis cannot reveal this difference. Specification-reduction testing treats measurement basis as an explicit design choice [40]. Observation incompleteness and sampling uncertainty are therefore separate limitations.

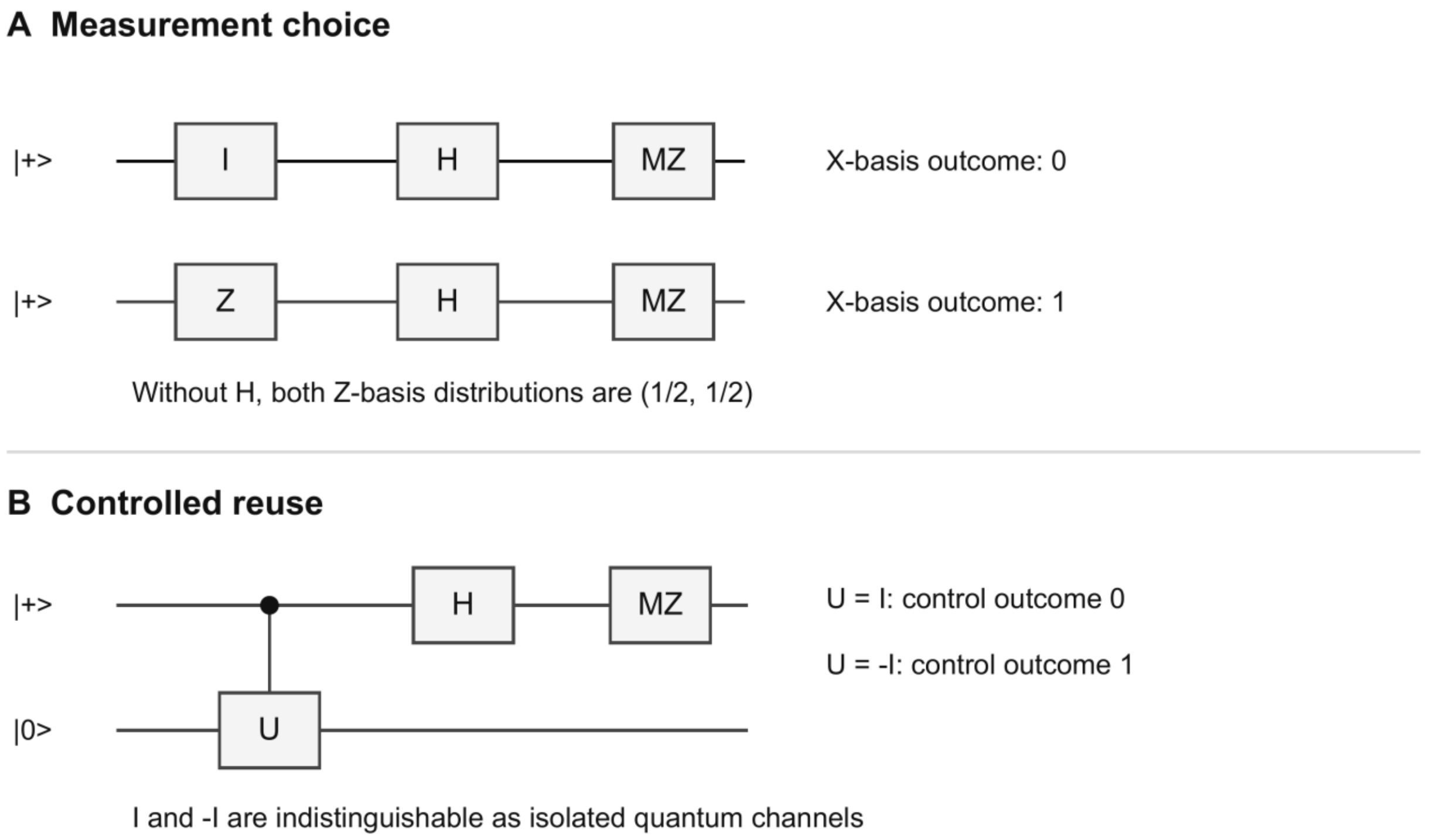

**Fig. 1** Analytical examples of observable differences. In panel A, I and Z applied to |+⟩ have identical Z-basis probabilities but different X-basis outcomes, obtained by applying H before measurement. In panel B, controlled-I and controlled-(−I) act differently on a control prepared in |+⟩ . These exact examples illustrate observation and composition. They are not new compiler experiments

### 3.4 Observation limits and complementary checks

The observation projection makes a basic limitation explicit. For a deterministic oracle o with all other inputs fixed, $\pi_o(O_1) = \pi_o(O_2)$ implies identical verdicts because the oracle receives identical inputs. With the same internal randomization distribution, a randomized oracle likewise has identical verdict distributions. This is an organizing observation, not a new verification theorem.

A fault affecting only a field omitted by $\pi_o$ cannot be detected by changing the decision algorithm while retaining that projection. For example, normalized gate semantics alone cannot reveal an isolated violation of a timing field. A checker that receives metadata may distinguish records that a gate-only adapter cannot. Capability claims must therefore identify the observations and normalization path, not only the checking algorithm.

Observation refinement and contract composition address different limits. An observation $\pi_2$ refines $\pi_1$ when $\pi_1$ can be reconstructed from $\pi_2$; any distinction available through $\pi_1$ is then also available in principle through $\pi_2$. Combining independently justified acceptance sets $A_1$ and $A_2$ yields $A_1 \cap A_2$. A smaller accepted set is useful only when both contracts apply: an incorrect additional requirement can reject valid outputs, and an expensive check can reduce tested coverage. Added detections, false alarms, and resource costs therefore require joint evaluation.

### 3.5 Compiler faults and changes in behavior

Fault, failure, and regression refer to different things and should not be used interchangeably. A fault is an implementation defect relative to a specification. A failure is an observed violation. A regression is a deterioration associated with a change from an earlier baseline on which the relevant obligation held. The distinction matters because two compilers may produce different but valid circuits, two seeds may produce different optimization outcomes, and a new release may intentionally change an interface. None of those differences is a failure unless an applicable contract is violated, and none is a regression unless the relationship to the baseline has been established.

A useful regression record needs more than a failing test. It should identify the baseline and candidate versions, the input and configuration that exposed the problem, the violated obligation, the observation supporting the verdict, and any competing explanations that remain unresolved. One underlying fault can generate many failing test cases, so failure counts and fault counts require separate identifiers. A crash on invalid input may reflect error handling rather than semantic corruption, while a timeout may arise from compiler performance or from the verifier's own resource limit. Keeping these cases separate is what allows the literature comparison that follows to connect each claim to the evidence that actually supports it.

## 4 Semantic verification methods and their boundaries

### 4.1 Verified compilers and validation of individual outputs

Verified construction proves properties of a transformation’s implementation within a formal model. Translation validation checks a specific source-output pair without requiring a verified compiler implementation. The former can establish preservation over an unbounded supported input family; the latter checks particular deployed outputs. Both depend on faithful representation of program semantics and interfaces.

VOQC proves optimizer transformations against formal SQIR semantics in Coq [6]. CertiQ reports mostly automated verification of 26 of 30 Qiskit passes, while Giallar reports automatic verification of 44 of 56 passes in the versions studied [44, 7]. These denominators concern supported passes, not the entire software stack. Their related tool lineages and any overlapping defects must be resolved before combining empirical counts.

VQO verifies translations from its restricted oracle language through OQASM to SQIR and separately uses property tests for selected oracle components [45]. QBlue proves a typed Hamiltonian-simulation compilation relation with backend-specific approximation bounds, including a gate-based path [46]. These studies widen the proof comparison: VQO distinguishes proved stages from tested components, while QBlue shows that the correct contract may be bounded approximation rather than exact unitary equality. Neither theorem covers an unrelated general-purpose transpiler or its deployed interface.

Earlier ReVerC proves preservation of Revs semantics and cleanup of temporary bits in reversible-circuit compilation [47]. ReQWIRE formalizes when an ancilla assertion permits safe discard and verifies a compiler for a restricted oracle language [48]. These results support a clean-ancilla obligation under their stated models. They do not make ancilla disposal sound in an unrelated pass merely because the output qubits agree.

RelOpt adds a recent cost-aware example: its reported QCore optimizer admits rewrites only after relational checks against user-specified contracts, including partial-trace, measurement, and approximate relations [28]. The published evaluation concerns that optimizer and its benchmarks; it does not establish preservation for arbitrary external compiler passes or a change-aware regression service.

Where the trusted boundary sits matters directly for regression analysis. A proof for a compiler pass can establish semantic preservation while leaving Python packaging, serialization, backend models, and integration code outside the proof. A change in any of those components can violate a compiler contract even when the mathematical transformation itself remains correct. The reverse is just as useful: a verified transformation can serve as a reference inside a testing system, as long as the implementation boundary is spelled out. The two forms of assurance are complementary rather than competing. Proofs restrict the class of possible transformation defects, and integration tests cover the obligations that fall outside those proofs.

### 4.2 Decision diagrams and algebraic representations

Explicit unitary matrices provide a direct comparison but require exponential space in the number of qubits. Decision diagrams exploit repeated structure in states and operators. Advanced checking combines reversibility, interleaved processing, and simulation-based counterexample search to control intermediate representations [8]. Performance nevertheless depends on circuit structure and the transformation relationship; a maximum qubit count on one benchmark does not establish general capacity.

Compilation-aware validation uses information about the transformation sequence rather than treating the source and output circuits as unrelated objects. Burgholzer, Raymond, and Wille combine reversibility with knowledge of the Qiskit compilation flow to guide equivalence checking and report verification of circuits containing tens of thousands of operations in seconds or less [49]. This provides a precedent for incorporating compiler context directly into semantic validation. The efficiency benefit, however, depends on how accurately that context reflects the transformation that was actually performed. The checker must still interpret source and output consistently when routing or another transformation changes the circuit representation.

Simulation-based verification makes a different trade-off. Liu and colleagues use specially designed input states intended to expose non-equivalence [50]. The relevant balance is between how many stimuli are used, how well those stimuli discriminate between circuits, and the cost of simulating each one. A bound on the number of required stimuli does not reduce the cost of an individual simulation. Such methods can produce counterexamples efficiently, but an agreement result is meaningful only in light of how the stimulus was constructed and the guarantee attached to that construction.

Numerical representation introduces another design decision. Accurate BDD-based operator manipulation stores supported amplitudes in a bit-sliced form, providing numerical robustness for that representation without claiming exact support for every arbitrary real-valued parameter [51]. Q-Sylvan explores a different point in the design space through parallel edge-valued decision diagrams, supporting both simulation and equivalence checking [52]. Parallel execution may reduce elapsed time, but it also increases processor demand. For a regression service operating under a budget, wall-clock time and total resource consumption therefore have to be reported separately.

ZX-calculus checking represents two circuits as diagrams and simplifies their relationship using sound graphical identities. Peham, Burgholzer, and Wille study this approach together with its complementarity to decision diagrams [9]. If the reduction closes the diagram, equivalence follows within the implemented semantics. If a heuristic reduction fails to close it, however, that outcome is not evidence of inequivalence. An integration must preserve an inconclusive result whenever the procedure cannot decide. A benchmark that maps every non-success outcome to a single failure flag loses exactly this distinction.

Weighted model counting provides another route from quantum semantics to classical reasoning. Mei and colleagues encode quantum-circuit equivalence as a weighted-counting problem using a Pauli-basis characterization [53]. Neither a small encoding nor a small number of counting calls implies polynomial-time solving. Huang and colleagues combine path-sum simplification with a weighted model-counting decision procedure, using algebraic reduction where it is effective and a complete residual check elsewhere within the formalized domain [54]. The example illustrates why several representations can be useful in one oracle portfolio: their strengths depend on instance structure, while the practical benefit still depends on implementation cost.

Manjushri represents local projections as discriminative circuit fingerprints in weighted binary decision diagrams [55]. Its comparison with model-counting approaches exposes structural trade-offs rather than a method that dominates every case. More generally, equivalence studies often use circuits whose transformation relationships are known in advance or whose changes are controlled mutations. Those benchmarks are appropriate for measuring solver capability. They do not by themselves tell us how often the solver will detect faults introduced by real compiler changes.

Solver evaluation and regression evaluation run on the same infrastructure, yet the unit of inference each draws from it is different.

The original AutoQ verifies circuit properties with symbolic tree-automata specifications [29]. Improved DD-based checking targets source/output circuit comparison and counterexample generation [32], whereas QuBEC adapts decision-diagram checking to circuits with quantum-error-correction embedding [31]. Their domains and representations differ; neither a circuit-property proof nor a QEC-specific comparison can be reported as a general deployed-compiler regression result.

### 4.3 Equivalence for parameterized partial and hybrid circuits

A July SLR cross-check identified two more observation strategies. QuaSi reports simulation-based transpilation checking with designed stimuli and uncompute checks that can localize discrepant qubits [56]. This gives useful counterexample evidence, but simulation agreement on selected stimuli is not a universal proof. A 2026 context-verified Toffoli pass instead checks each proposed relative-phase or approximate substitution in its circuit context [57]. Its preprint reports 66 non-equivalent library rewrites without the context gate and corruption of six of twelve benchmarks by count-greedy substitution; these are author-reported results for that study, not a cross-method detection rate. The pair shows why a resource objective needs an explicit semantic admission rule.

Parameterized circuits require an explicit quantifier: correctness at one binding differs from correctness over every permitted binding. Peham et al. combine parameter-aware checking approaches, while Hong et al. develop symbolic reasoning [4, 58]. Comparing their guarantees requires the supported expressions and gates, parameter domain, and completeness argument. Finite sampled agreement establishes a universal identity only when an applicable theorem justifies that reduction.

A newer restricted-domain result checks Clifford layers interleaved with shared single-qubit layers for all choices of those shared unitaries [30]. That all-choice guarantee depends on the shared-layer structure; it is not a general theorem for every parameterized circuit or dynamic program.

Partial equivalence restricts the permitted inputs or observed outputs. Chen et al. formalize agreement of selected measurement probabilities over allowed initialized inputs [3]. Such a relation can match a contract involving ancillas or discarded outputs, for which full-unitary equality would be inappropriate. Differences outside the declared observations remain unconstrained. Its adequacy therefore depends on intended use rather than an unconditional ranking against full equivalence.

Dynamic circuits bring measurements and classical control into the picture, and a plain unitary representation is not enough for them. Ricciardi and colleagues extend unitary checking with deferred-measurement transformations and separation and projection techniques, reaching hybrid comparisons that involve initialization and discards [41]. The transformation comes with its own prerequisites, and it does not turn every non-unitary program into a routine unitary check. A regression harness therefore has to record the lifted problem and its assumptions explicitly, particularly when a unitary source is being compared against a hybrid implementation, so that a failure in the lifting can be told apart from a genuine counterexample to the original program's behaviour.

Non-unitary checking already has compilation-oriented precedents. Burgholzer and Wille describe reset elimination and deferred measurement for a lifted unitary comparison, and a separate simulation scheme comparing output distributions for fixed inputs [59, Secs. 3–5]. Reset elimination adds qubits, while selected-input distribution agreement establishes a narrower relation. These alternatives support different contracts; neither should be reported simply as unrestricted dynamic-circuit equivalence.

Approximate and noisy equivalence bring in a metric and an error policy together. Hong and colleagues use a channel-fidelity formulation built on tensor networks. Huang and colleagues instead look at approximation under low-noise conditions [42, 43]. A fidelity threshold, a norm bound, and agreement of sampled distributions are three distinct relations, and lumping them under a single unlabeled “equivalent” field erases the difference. Nor should a deliberately approximate optimization be marked wrong just for breaking exact equality. The expected tolerance has to be set independently of whatever discrepancy turns up, with numerical error, algorithmic approximation, and hardware noise held apart from one another.

Automata-based verification adds one more specification mechanism. AutoQ 2.0 checks Hoare-style properties over automata representations of state sets, with user-provided loop invariants and built-in support for measurement and classical control, handling a weak-measurement version of Grover's search at 100 qubits in about 20 minutes and verifying repeat-until-success algorithms with short reported runtimes [60]. It delivers partial-correctness evidence for the specification it was given, not a general termination proof and not an open-ended compiler-equivalence procedure. Compiler use requires a specification expressing the relevant preservation obligation.

Circuit-size parameterization differs from symbolic gate angles. AutoQ-Para represents supported size-indexed families using synchronized weighted tree automata [61]. Its functional equivalence and inclusion problems are decidable, whereas language equivalence and inclusion are undecidable [61, Theorems 5.6 and 5.9]. The implementation restricts coefficients to algebraically representable complex numbers. Five case studies illustrate supported families [61, Sec. 4.6, Table 1]; they do not verify an arbitrary compiler. These qualifications are essential when comparing size-parameterized verification with angle-parameterized checking.

FeynmanDD uses decision-diagram counting over path-integral representations [62], while intermediary matrix product operators exploit reversibility during tensor contraction [63]. Hong et al. describe dynamic-circuit equivalence using ensembles of linear operators [64]. These mechanisms enter the same evidence matrix as other semantic representations, with access depth stated individually. Qafny proves supplied program specifications through a typed proof system and a translation to Dafny [65, Sec. 4]. Its later QafnyC implementation separates verification translation from the execution path through SQIR and OpenQASM [66, Sec. 1, Fig. 1]. Program-level verification is relevant when its specification expresses the compiler obligation, but a program proof does not automatically certify every compilation stage.

The restriction to parameter-free checking is not unavoidable. Ross and Wesley reduce equivalence for supported rational-linear parameterized rotations to finitely many parameter-free checks, with a probabilistic alternative when exhaustive combinations are costly [24, Secs. 3–6]. Their guarantee relies on the derived cutoff and arithmetic assumptions; arbitrary finite sampling does not inherit it. QUESO uses polynomial identity testing to verify synthesized symbolic rewrites with a probabilistic guarantee [25]. It therefore occupies a different evidence category from both a deterministic pass proof and uncalibrated random testing.

Shallow-circuit structure provides another scalability boundary. Yu and colleagues develop projection-based checks for fixed-input and arbitrary-input equivalence whose complexity is linear in qubit count for fixed depth [26, Secs. 4–6]. Increasing depth is not covered by that fixed-depth scalability claim. Local projections here form a complete representation under stated structural assumptions; this should not be confused with selecting a few arbitrary measurements of a general circuit.

### 4.4 Learning inside a verifier and learning a verdict

Machine learning can shape verification without ever taking over the semantic decision rule. Al-Rousan and colleagues study reinforcement learning for tensor-contraction ordering across 110 circuits from five circuit families at 2 to 23 qubits, reporting contraction roughly two orders of magnitude faster than hypergraph-based heuristics on Bernstein–Vazirani instances, where the learned part only decides how a semantic computation is carried out [67]. ZXNet puts learning somewhere else entirely, using a graph-neural approach to classify equivalence [68]. The two placements carry different trust obligations. A learned search policy can keep the underlying checker sound as long as it only reorders sound operations and calls an unsuccessful search inconclusive. A predicted equivalence label requires separate validation before it can justify accepting a compiler output.

Classification accuracy reported for a learned model is not a proof guarantee. It depends on the training and test distributions, on how equivalent and inequivalent examples were constructed, and on the overlap among those families. Learning can still be useful for triage, for example by directing suspicious cases to an authoritative checker or by predicting which representation is likely to finish quickly. In that setting, evaluation has to cover the full pipeline, including false negatives introduced by triage. The role of learning should also be explicit in the architecture so that a speed improvement does not silently weaken the evidence required for acceptance.

Table 4 places the semantic relations beside their operational limits. Decision-diagram storage and state simulation may still grow exponentially. Symbolic simplification remains sensitive to circuit structure. And proof-based assurance shifts effort into formalization and trusted integration rather than removing it. For that reason, maximum qubit counts reported on different workloads do not form a common scalability metric. Supporting Information 2 keeps each evaluation setting separate precisely so that method-level distinctions are not confused with workload-specific limits.

**Table 4** Semantic assurance relations and operational limits

| Approach | Relation and scope | Result that can be accepted | Resource or integration boundary |
|---|---|---|---|
| VOQC [6] | Formalized optimizer functions over supported SQIR circuits | Preservation theorem for the stated pass domain | Proof development; extraction and external interfaces |
| Giallar [7] | Symbolic verification of supported compiler passes | Pass preservation under supported rules and loop templates | Unsupported pulse, solver, randomized and approximate passes in study |
| Decision diagrams [8, 49] | Supported unitary operators with interpreted qubit order | Completed equivalence decision under numeric assumptions | Intermediate representation can grow exponentially |
| Designed stimuli [50] | Responses to method-specific distinguishing states | Detection or assurance justified by stimulus construction | State simulation cost remains separate from number of stimuli |
| Exact BDD encoding [51] | Algebraically encoded unitary entries | Exact result in the supported number domain | Representation growth and gate-domain restrictions |
| ZX checking [9] | Sound rewriting of a circuit comparison | Identity derivation; complete Clifford case discussed | General simplification is incomplete |
| Symbolic parameters [4, 58] | Supported symbolic parameter relations | Universal result only under the method's assumptions | Numeric samples alone do not establish an identity |
| Partial checking [3] | Selected probabilities for all allowed initialized inputs | The stated partial observational equivalence | Does not preserve every unmeasured quantum property |
| Hybrid lifting [41] | Deferred-measurement reduction and interface-specific relation | Result when lifting and projection assumptions hold | Added width, supported control constructs, preprint evidence |
| Model counting and path sums [53, 54] | Encoded semantic identities | Result under the encoding and decision procedure | Encoding size and solving complexity |
| Approximate checking [42, 43] | Specified process metric and tolerance | Bound or decision for that metric and domain | Noise model and numerical error cannot be conflated |

| Approach | Relation and scope | Result that can be accepted | Resource or integration boundary |
|---|---|---|---|
| Automata specifications [60, 61] | State-set contracts or supported circuit-size families | Verification under the supplied specification | Specification construction and supported program syntax |
| Learned classification [68] | Prediction on represented circuit features | Triage evidence rather than an unqualified proof | Distribution shift and false acceptance require independent checking |

BDD: binary decision diagram; SQIR: Small Quantum Intermediate Representation; VOQC: Verified Optimizer for Quantum Circuits; ZX: graphical calculus using Z- and X-type spiders.

## 5 Testing methods and demonstrated fault evidence

### 5.1 Differential and metamorphic testing

Differential testing compares implementations or execution paths that are expected to agree. QDiff exercises quantum-software stacks by generating circuits, applying equivalent transformations, and comparing platforms. It also uses circuit and device information to avoid executions judged unsuitable for hardware testing [10]. Across six seed algorithms, the study produced 14,799 transformed programs and reported four software crashes, together with two potential explanations for hardware divergence. The hardware observations are not two additional confirmed compiler defects. The filtering step therefore consumes part of the testing effort before execution begins. Chronological compiler regression uses a related idea for a different objective: directing a fixed budget toward deterioration associated with specific changes.

Agreement across implementations is informative only when the conditions for agreement are justified. Several platforms may share a dependency, the same interpretation error, or a common bug. Conversely, a discrepancy may arise from incompatible conventions or noise assumptions rather than from a defect. A convincing differential result therefore needs three pieces: a relation that should hold, normalization of the semantics being compared, and diagnosis of the observed disagreement. Treating one implementation as a golden oracle without examining its trust boundary does not remove the oracle problem. It simply relocates it.

MorphQ creates related Qiskit programs through metamorphic relations and compares their outcomes [11]. On Qiskit 0.19.1, it compared 8,360 program pairs in 48 hours and filed 13 bug reports, nine of which were confirmed at publication. MorphQ++ later replicated and extended the study, on Qiskit 0.44.1 with six new and two extended transformations, using three 48-hour campaigns [69]. Their generation strategies are relevant here, but so is the failure channel: the reported bugs were crashes rather than the output discrepancies that the method was designed to compare. That does not invalidate the output oracle. It does mean that these results cannot be used as evidence of broad semantic-regression detection. Method-capability summaries therefore need to identify which failure channel produced each reported finding.

Quantum equivalence modulo inputs makes the input restriction explicit. QEMI constructs program variants expected to agree on selected inputs and uses adaptive stopping while evaluating their outputs, so generation cost and measurement expenditure are considered together [12]. Across Qiskit 2.0.0–2.1.0, a Cirq development build, and Q# 1.17, the study reported 11 crashes and one distribution discrepancy. Agreement on selected inputs is weaker than universal circuit equivalence, but it can still support effective bug discovery without requiring a complete output specification. QSPE addresses a related problem using skeletal quantum programs and statevector-based output validation [70]. Statevector comparisons on selected inputs remove shot noise in that setting, yet they still do not constitute a full-unitary proof. Counts of generated variants or approved reports must also be deduplicated before they are interpreted as counts of distinct compiler defects.

QASMInfer provides an independently derived outcome-distribution oracle for supported OpenQASM programs [39]. The published abstract and artifact description indicate support for dynamic constructs and a mechanized core. Because the full formal development and evaluation were not assessed, this review uses the source only to establish the oracle's stated role. Its reported bug totals are excluded from Table 5 and do not support a comparative effectiveness claim.

IterTestQ repeatedly imports, transforms, and exports assembly programs across platforms, combining crash monitoring with QCEC comparisons for selected program pairs [71]. Its evaluation reports 23 platform issues, 17 of them confirmed or fixed, across five platforms, out of a bug-finding campaign that ran several times over three months. This is a precedent for combining interface exploration with semantic checking. The incremental detection value and reliability of particular combinations still require evaluation on controlled change events. The author manuscript also reports defects in the checker itself, which is an important reminder that an oracle deployed in practice is software and requires assurance of its own.

### 5.2 Properties and assertions for targeted generation

The SLR cross-check also found Fuzz4All's Qiskit target [72]. Its published multi-system study uses generated Python inputs and a crash-oriented oracle for Qiskit transpilation; the paper's aggregate bug total spans nine systems and cannot be attributed to the quantum compiler alone. Crash detection therefore complements, but does not replace, a semantic miscompilation oracle.

Property-based testing replaces a complete expected output with a relation that should hold over a family of inputs. QuCheck follows this pattern by providing extensible generators, assertions, and preconditions for Qiskit programs [14]. The inspected 2025 preprint evaluation associated with this publication covers five programs at 2 to 10 qubits and varies properties, inputs, and shots independently. Those dimensions represent different kinds of testing effort. Additional properties can exercise different behaviours, whereas additional shots mainly strengthen evidence for observables that have already been selected. Applying the same idea to compiler testing requires an extra justification: the property must remain invariant under the transformation being tested and remain valid for the relevant backend and parameter domain.

Specification reduction with mixed-basis projective measurements changes both observation cost and fault sensitivity [40, Secs. 3–5]. It therefore belongs in the same comparison as property-based testing, despite being evaluated on quantum programs rather than compiler-change events. Fixed-point amplitude amplification offers a different statistical testing mechanism [73]. Both require an explicit account of the specification and quantum access available to the test; neither is a general replacement for a source-output equivalence check.

Black-box relation testing provides a further observation model. Long and Zhao treat equivalence, identity, and unitarity of quantum operations with classical parameters fixed [74, Sec. 3]. Their comparison algorithms require suitable quantum inputs and access to output states; those requirements differ from reading classical result counts alone. QuCAT varies combinatorial input strength [75], while multi-subroutine testing addresses unit and integration relations [76]. These studies inform generation and composition at a compiler boundary, but require an independently justified preservation property before a failure can be attributed to compilation.

Concolic testing for quantum compilers combines symbolic reasoning about compiler decisions with concrete execution. In the reported work, it targets selected Qiskit compiler decisions and QASMBench circuits in order to reach difficult paths and optimization behaviour that simpler generators may miss [77]. Coverage of the modeled decisions is evidence that the generator reaches those conditions, but it is not coverage of every compiler contract. Missing an optimization may violate a quality objective without violating semantic correctness, because the resulting circuit can remain functionally correct even when a transformation opportunity is missed. Concolic testing therefore illustrates a broader point: relevant compiler observations extend beyond output equivalence.

Static analysis can flag unsupported quantum-program patterns before a candidate ever reaches compilation or simulation, and LintQ is a static-analysis framework for Qiskit that detects patterns tied to program errors in just this way, evaluated on 7,568 real-world Qiskit programs at 91.0 percent precision in its default configuration [78]. Analysis like this helps validate test inputs and flag suspicious interfaces, yet a warning on its own is not evidence that a compiler change caused a regression. Diagnosis gets further support from the assertions and debugging methods Rosa and Santiago survey, intermediate-stage observations among them [20]. Turning any of those observations into a regression oracle, though, means specifying the expected condition and where it applies independently of the failure being investigated.

### 5.3 Regression diagnosis and mutation evidence

Delta debugging has been applied to property-based quantum-program regression testing, including an evaluation using injected faults in three algorithms [79]. Failure reduction supports reproduction and diagnosis, but does not establish the effectiveness of compiler-level test selection. That evaluation additionally needs explicit obligations, eligible actions, a scheduling policy, and independently attributed change events.

Mutation testing introduces controlled faults and asks whether the test system detects them. Muskit, validated on four quantum programs, and later empirical work that generated 723,079 mutants from 382 circuits across 28 algorithms provide both tooling and evidence for this approach [80, 81]. Controlled mutations are useful for isolating observation channels because a mutation can be designed to alter phase, structure, or metadata while leaving other aspects unchanged. The latter study reported 48% survival under its wrong-output and probability oracles, using an all-zero input and 100,000-shot simulator setting; that figure is not a compiler-regression miss rate. A surviving mutant is not necessarily a missed fault, however. It may be equivalent, violate an input assumption, or never reach the observation point. A mutation score therefore needs an explicit denominator and a stated policy for equivalence it could not resolve.

Transpiler-based mutant-equivalence detection compares normalized compilation outputs [82]. Using the compiler under assessment to certify its own benchmark labels can introduce circular evidence. Labels used for compiler evaluation therefore require an independent checker, construction-based proof, or separately trusted transformation where possible. Related variants of one mutation must also be distinguished from independent observations of real fault prevalence.

Bugs4Q contains 36 manually validated Qiskit bugs from Terra, Aer, Ignis, and Aqua and was assembled for controlled testing and debugging studies [83]. Its real-world origin is valuable, but the benchmark still has to match the claim being evaluated. Application-level quantum bugs do not automatically form a compiler-regression benchmark. The broader platform study raises the same issue. Of 223 bugs collected from 18 platforms, only 39.9% were quantum-specific, and the dataset spans several component types [84]. Reusing either benchmark therefore requires explicit filtering, provenance, and an explanation of how the retained defects relate to the compiler contracts under evaluation.

QuteFuzz extends program generation into nested control flow and subcircuits, targeting configurations that simpler gate-sequence generators may not reach [13]. Across Qiskit, Cirq, and Pytket, its 17 reported failures cover both compiler and simulator components. QEMI later compares itself with QuteFuzz and earlier tools on specific Qiskit versions [12]. Because those approaches are evaluated within the same study, the comparison provides stronger evidence for those particular settings than a comparison of totals drawn from separate campaigns. It still does not establish a version-independent ordering of the methods.

Table 5 retains source-specific evaluation denominators. QASMInfer is excluded from the numerical comparison because only its published abstract and artifact description were assessed. QuCheck findings are restricted to the inspected earlier manuscript and descriptive conclusions consistent with the online journal abstract; changed cross-version performance percentages are not imported.

**Table 5** Evaluation settings and limits of reported findings

| Study | Assessed evaluation setting | Reported evidence | Inference supported and excluded |
|---|---|---|---|
| Giallar [7] | 44 of 56 passes in 13 reported Qiskit versions | Three confirmed bugs; 12 unsupported passes | Supported pass verification, not whole-stack certification |
| QDiff [10] | Six seed algorithms; 14799 transformed programs | Four software crashes; two potential explanations for hardware divergence | Stack evidence; potential explanations are not confirmed compiler defects |
| MorphQ [11] | Qiskit 0.19.1; 8360 pairs in 48 hours | 13 bug reports, nine confirmed at publication | Crash discovery; warning and report counts are distinct |
| QuteFuzz [13] | Qiskit, Cirq, Pytket; multiple generated-circuit runs | 17 reports spanning compilers and simulators | Richer constructs reveal defects; total is not compiler-only |
| QEMI [12] | Qiskit 2.0.0–2.1.0, Cirq development build and Q# 1.17 | 11 crash reports and one distribution discrepancy | Includes a rediscovery; not 12 independent forward-regression events |
| IterTestQ [71] | Five platforms; multi-run campaign | 23 issues; 17 confirmed or fixed | Interoperability evidence; not prospective regression recall |
| QuCheck [14] | 2025 preprint: five programs, 2–10 qubits | Property, input and shot configurations affect detection | Application mutation evidence; no compiler-regression recall inferred |
| Concolic testing [77] | Selected Qiskit decisions and QASMBench circuits | Decision, boundary and outcome coverage | Coverage denominator is the studied decisions |
| Platform bug study [84] | 223 bugs from 18 platforms | 39.9% quantum-specific in that corpus | Descriptive corpus finding; not field-wide compiler prevalence |
| Benchpress [15] | Seven SDKs; over 1000 tests in author manuscript | Processing workloads up to 930 qubits | Processing scale, not full-state verification at that scale |

SDK: software development kit; QASMBench: benchmark suite of quantum assembly programs.

# 6 Target constraints and interpretation of compiled circuits

## 6.1 Circuit quality and compilation performance

Compiler output quality has several dimensions, including depth, two-qubit count, connectivity compliance, scheduling duration, and hardware-informed error estimates. Improving one metric can worsen another. A deterioration is a regression only relative to a declared objective or monitoring criterion; otherwise it is a reported trade-off rather than evidence of a defect.

Benchpress evaluates quantum-software performance and functionality across seven SDKs, with more than 1,000 tests and processing workloads reaching 930 qubits [15]. Open QBench supplies another platform-benchmarking framework [85]. Such studies support workload and measurement design, but identifying a regression additionally requires a controlled change and separation of unexpected deterioration from changes in optimization goals, backend descriptions, or supported workloads.

Processing scale and semantic-checking scale should be reported separately. A large compilation workload may support target-validity and resource measurements even when full-state validation is infeasible. Smaller instances may support stronger semantic comparisons. This heterogeneous coverage is useful when the obligation assessed at each scale is explicit; a performance measurement alone does not establish semantic preservation.

Performance comparisons require control of classical execution variability. Scheduling, thermal conditions, caching, garbage collection, thread counts, and dependency initialization can affect measured runtime. Repeated, randomized, or interleaved baseline and candidate runs can reduce confounding. Absolute and relative tolerances should be chosen before inspecting outcomes and justified by the service objective and workload.

### 6.2 Metadata and backend contracts

Metadata determines how a compiled circuit is used. Layout maps logical to physical qubits, parameter mappings associate symbolic values with inputs, and result mappings connect measured bits to application outputs. These records may be incorrect even when the gate sequence is well formed. Their validation requires an independently justified relation; equality with a previous release is insufficient because correct compilation may change the mapping.

Consider an unchanged physical circuit with a corrupted delivered output permutation. A checker that substitutes an independently reconstructed correct permutation can accept the gates without checking the delivered record. An adapter that uses the delivered permutation, or a separate mapping check, can expose the violation. Section 7.4 formalizes this distinction. Evaluation should compare additional valid detections under explicit adapter and normalization assumptions.

Backend validity is version-dependent. Supported operations, connectivity, and timing constraints derive from the selected backend description. To attribute a failure to a compiler change, a comparison should hold the backend snapshot fixed or explicitly include its change in the event definition. Failure to meet a target constraint differs from a quality trade-off between mappings that both satisfy it.

A recent preprint by the present authors analyzes compiler fixes whose manifestations are not captured by a specified output-equivalence screen [34]. It is relevant empirical context for interface obligations, but its fix classifications are not a prospective regression-detection evaluation. The screen's normalization and metadata policy are essential: a layout-aware checker that validates the delivered interpretation has a different observation boundary. The worked example below establishes that distinction analytically without relying on the preprint's numerical estimates.

### 6.3 Reproducibility and provenance

Two recent pipeline studies expose additional boundaries. EmuPlat describes transformation validation across circuit transpilation, pulse compilation and simulation, but its preprint’s small Bell and QFT demonstrations do not establish historical regression effectiveness [86]. An OpenQASM 3-to-CUDA-Q transpiler reports tests of feedforward, reset and parameterized circuits [87]; its conference-linked preprint illustrates why a dynamic-circuit assurance record should retain both control-flow outcomes and parameter bindings. Neither study evaluates version-triggered test selection.

Reproducibility may require repeatability under fixed seeds, semantic consistency, or stability of quality distributions. Byte-identical serialization is inappropriate when ordering, names, or formatting may legitimately vary. The oracle should state its normalization and required level of stability. Seeds, dependency versions, backend snapshots, and process-level randomness belong in the observation record.

Compiler tracking by Stirbu and Meijer–van de Griend records compilation stages, pass information, backend characteristics, configuration, and results [88]. This supports reconstruction of compilation decisions. Linking those records to versioned contracts can support regression adjudication, although trace completeness does not establish correctness.

Provenance makes a comparison auditable but does not validate its specification. A complete trace may reproduce a faulty compilation or an inappropriate comparison. Auditing empirical comparisons similarly requires attention to the scope and conditions of evidence [89]. A regression record should therefore preserve the contract, normalization, exclusions, timeouts, and unresolved outcomes alongside successful outputs.

Peltonen and colleagues provide a concrete baseline-relative workflow that records compiled artifacts, metrics, and provenance and emits configurable CI warnings or failures [27, Secs. IV–VI]. Its demonstration changes seeds or the transpilation toolchain. Structural drift is explicitly a monitoring signal; it does not by itself establish a semantic defect. This precedent narrows the research gap to justified obligations, change attribution, and the additional detection value of checks, rather than the invention of compiler monitoring.

### 6.4 Documented cases and their assurance implications

Four cases connect the taxonomy above to concrete evidence. They were selected to illustrate different assurance decisions, not to estimate defect frequency. Each case's sources, inspection dates, and inference limits are recorded in Supporting Information 3. Cases A and B are drawn from research publications, Case C combines a published comparison with an explanatory conjecture that we label as such, and Case D comes from an engineering issue report, where the issue being closed is not read as evidence that any particular repair was actually made.

Case A concerns an invalid precondition for gate merging. In its analysis of optimize_1q_gate, Giallar found that the merge logic could overlook classical or quantum conditions attached to gates [7, Sec. 7.1 and Fig. 8]. A rewrite that is valid for two unconditional rotations does not necessarily preserve a conditional operation. The authors report adding checks for these conditions to the verified implementation. From an assurance perspective, a generator restricted to unconditional gates cannot exercise this boundary no matter how widely it varies rotation angles. The algebraic rewrite and the guard under which it executes have to be tested together. This case links semantic correctness with transformation applicability. It does not show that one verifier subsumes all forms of testing.

Case B concerns a parameter-order error in ConsolidateBlocks. QuteFuzz reports that a refactoring exchanged two parameters and that existing unit tests used equal values, so the swap remained invisible [13, Secs. 3 and 4.1.1]. The case gives a direct reason to include unequal and boundary parameter bindings in executable tests. Where its assumptions hold, a symbolic identity check can cover a broader parameter domain. Sampled tests remain useful for compiler and adapter code outside the formalized boundary. The two mechanisms therefore strengthen different parts of assurance: one through representation strength and the other through input discrimination. The paper also reports simulator defects, so its overall bug count is not a compiler-only total.

Case C concerns agreement on failure. In the QEMI comparison, QuteFuzz exposed Qiskit issue 13165 while QEMI did not. The authors conjecture that identical crashes in the original program and its variant may be treated as consistent by the comparison strategy [12, Sec. 4.3]. That conjecture has not been independently reproduced as a root-cause finding, but it exposes a clear contract-composition issue. Two executions can agree with one another and still both violate a valid-input termination requirement. A separately justified validity check can therefore complement a relational oracle without changing the oracle's semantic relation. Unsupported inputs must still be excluded before a crash is classified as a defect.

Case D concerns a phase requirement that depends on reuse context. Qiskit issue 16377 reports that CommutativeCancellation in version 2.4.1 adds a phase of $\pi$ when a sequence of P/U1-type rotations accumulates to $2\pi$ [90]. The report supplies identity and $-I$ matrices, requests strict operator equality, and itself notes the consequence under coherent control. Those are source-reported statements. Our contribution is the explicit separation of acceptance relations: I and $-I$ define the same isolated channel, but controlled-I and controlled-($-I$) differ. The controlled-use distinction is therefore not claimed as a new observation. The report's proposed cause and remedy are not an independently verified repair; closure alone identifies neither a repair commit nor the first corrected release.

**Table 6** Documented cases linked to method selection

| Case and evidence | Obligation and detecting observation | Alternative-check limitation | Consequence for Table 8 |
|---|---|---|---|
| A Giallar [7], Sec. 7.1, Fig. 8; reported defect | Preserve guarded operations; inspect conditions and conditional behavior | Unconditional-only inputs cannot exercise the guard | Check pass prerequisites and generate conditional cases |
| B QuteFuzz [13], Sec. 3 and 4.1.1; reported defect | Preserve parameter association; use unequal bindings or supported symbolic reasoning | Equal bindings conceal a swap | Pair parameter reasoning with adapter and binding tests |
| C QEMI [12], Sec. 4.3; reported miss, conjectured cause | Eligible input must terminate successfully; observe status separately from relational agreement | Matching crashes may agree yet violate validity; explanation remains conjectural | Add an independent validity obligation before interpreting agreement |
| D Issue 16377 [90]; engineering report | Preserve exact phase if the API or controlled reuse requires it; observe operator phase or controlled interference | Isolated channel equality intentionally ignores global phase | Declare phase contract before selecting or composing checks |

API: application programming interface.

## 6.5 Cross-study findings

Five findings organize the synthesis. Their evidence types matter: a theorem establishes a scoped guarantee, an evaluation reports behavior in its setting, and a counterexample demonstrates a logical limitation. None alone measures the benefit of a deployed portfolio. Table 7 consolidates the chain from source evidence to inference; the source-level ledger and qualifications are in Supporting Information 2.

**Table 7** Evidence supporting the cross-study findings

| Finding | Source evidence and type | Synthesis inference | Claim not established |
|---|---|---|---|
| F1 Assurance has several dimensions | Partial and parameterized relations [3, 4]; cutoff theorem [24]; formal results | Input states, parameters, observables, and interface boundaries must be stated separately | One universal ordering of assurance methods |
| F2 Generation and observations address different misses | Conditional-merge and parameter-order cases [7, 13]; reported defects | A suitable oracle still needs a triggering input; a triggering input still needs a discriminating observation | Measured marginal gain from combining these tools |
| F3 Deployed adapters constrain mathematical guarantees | Dynamic-circuit comparison [5], Sec. 4, Table 2; empirical outcomes | Representation conversion and outcome categories must be audited before interpreting success rates | A general soundness-error rate for equivalence checking |
| F4 Evaluation denominators differ | Benchpress processing workloads [15]; QuCheck property experiments [14]; empirical findings | Processing scale, property sensitivity, and distinct fault detection answer different questions | Pooled effectiveness or operational reliability |

| Finding | Source evidence and type | Synthesis inference | Claim not established |
|---|---|---|---|
| F5 Complementarity is contract-dependent | QEMI miss and conjecture [12]; phase report [90]; layout definitions [91]; analytical example in Sec. 7.4 | Add observations only where they test an applicable obligation and have an explicit interpretation | Validated optimal portfolio or guaranteed independence |

F1 is grounded in the definitions and guarantees of the compared methods. F2 combines documented failures with a logical distinction between reaching a fault and observing its effect. F3 requires care: the unsuccessful outcomes in the dynamic-circuit study include timeouts and undecidable outcomes as well as inequivalence verdicts. They cannot all be counted as incorrect decisions. F4 follows from different evaluation objects, not from a statistical comparison across studies. F5 is partly analytical: the shared-crash explanation is conjectural, whereas the isolated-versus-controlled phase distinction follows directly from the two operators.

The evidence therefore supports a conditional selection procedure: define the obligation, choose a relation, check input and representation applicability, and identify the observation needed for a verdict. Whether that procedure improves detection at a fixed cost remains an empirical question. The recent drift workflow [27] supplies implementation precedent for monitoring, while the compiler-fix preprint [34] supplies related defect evidence; neither validates the full procedure proposed here.

## 7 Selecting and combining assurance methods

### 7.1 Select the claim before selecting the tool

Method selection begins by specifying whether the claim concerns a transformation implementation, a compiled artifact, a program property, or a compiler revision. This fixes the input domain, parameters, observations, and reuse context. Table 8 translates the comparisons in Sections 3–6 into conditional choices; it is a synthesis guide, not an empirically established ranking.

**Table 8** Method selection for compiler assurance

| Assurance task | Suitable starting point | Necessary complementary evidence | Acceptance boundary |
|---|---|---|---|
| Prove a supported optimization pass | Mechanized or symbolic pass verification [6, 7] | Tests of parsing, extraction and integration | The formalized pass and its prerequisites |
| Validate one compiled unitary circuit | Equivalence checking with reconstructed layout [8, 9] | Target validity and adapter checks | The compared pair and declared phase convention |
| Validate symbolic gate bindings | Parameter-aware reasoning [4, 58] | Unequal and boundary bindings in integration tests | Supported parameter expressions and quantification |
| Validate measurement and control flow | Hybrid relations or specified program properties [41, 60] | Branch records and applicability tests | Stated process or property, including input restrictions |
| Explore transformations across representations | Cross-platform generation and equivalence checks [71] | Independent input validity and failure diagnosis | Supported transformations and observed comparisons |
| Test when full semantics is too expensive | Properties, metamorphic relations or selected observations [14, 12] | Explicit blind spots and error policy | Tested properties, inputs and observables |

| Assurance task | Suitable starting point | Necessary complementary evidence | Acceptance boundary |
|---|---|---|---|
| Detect deterioration across revisions | Reproducer suite with matched versions and targets [92, 79] | Change attribution and resource accounting | Independently justified version-to-version failures |

For an unsupported instance, an alternative representation may preserve the original assurance claim. A fallback to selected-input testing changes that claim and must be reported as restricted coverage. Computational-basis sampling does not inherit an all-input equivalence guarantee because both procedures return a passing result. If an essential obligation remains unresolved when the budget is exhausted, an overall assurance pass is unjustified.

### 7.2 Combine observations without counting agreement twice

A portfolio may combine valid-input termination, semantic preservation, output interpretation, and target compliance, provided that each check has its own independently justified acceptance rule. Cases A and B show why discriminating conditions and parameter bindings have to be generated. Cases C and D make the complementary point: a stronger acceptance condition is useful only where the interface actually requires it. Elsewhere it consumes resources without strengthening the relevant claim. Shared parsers, layout reconstruction, and numerical libraries also need to be recorded because a common defect in this infrastructure can affect otherwise different checking algorithms in the same way.

Complementarity should be evaluated on the same eligible workloads and against adjudicated faults. The report should show what each oracle detected, where detections overlap, which additional false alarms appeared, which outcomes remained unresolved, and what the total cost was. Removing one observation at a time, under comparable budgets, helps isolate its incremental detection value. A second checker that confirms a failure can be valuable for diagnosis, but it does not create a second independent fault. Nor do two checkers provide independent corroboration if both rely on the same faulty input conversion.

### 7.3 Allocate effort after preserving the evidence standard

Classical regression research already separates minimization, selection, and prioritization, and works through cost-aware, historical, and adaptive strategies for each [92, 93, 94, 95, 96], while reinforcement learning, bandit, and predictive approaches offer scheduling mechanisms of their own [97, 98, 99]. On the quantum side, studies already vary input selection, filtering, and measurement effort [100, 10, 12, 14]. None of it converges on one preferred scheduler for compiler regressions. It does make the case for budgeted assurance.

A testing action consists of a circuit, target, configuration, and eligible oracle. Shared compilation should be charged once, with incremental conversion, checking, sampling, and policy costs reported separately. Elapsed time, processor use, and peak memory capture different resource constraints. Mandatory obligations define the minimum acceptance evidence; optional checks can then be ordered by cost, diversity, or historical evidence. Supporting Information 4 gives a possible allocation formulation for future evaluation.

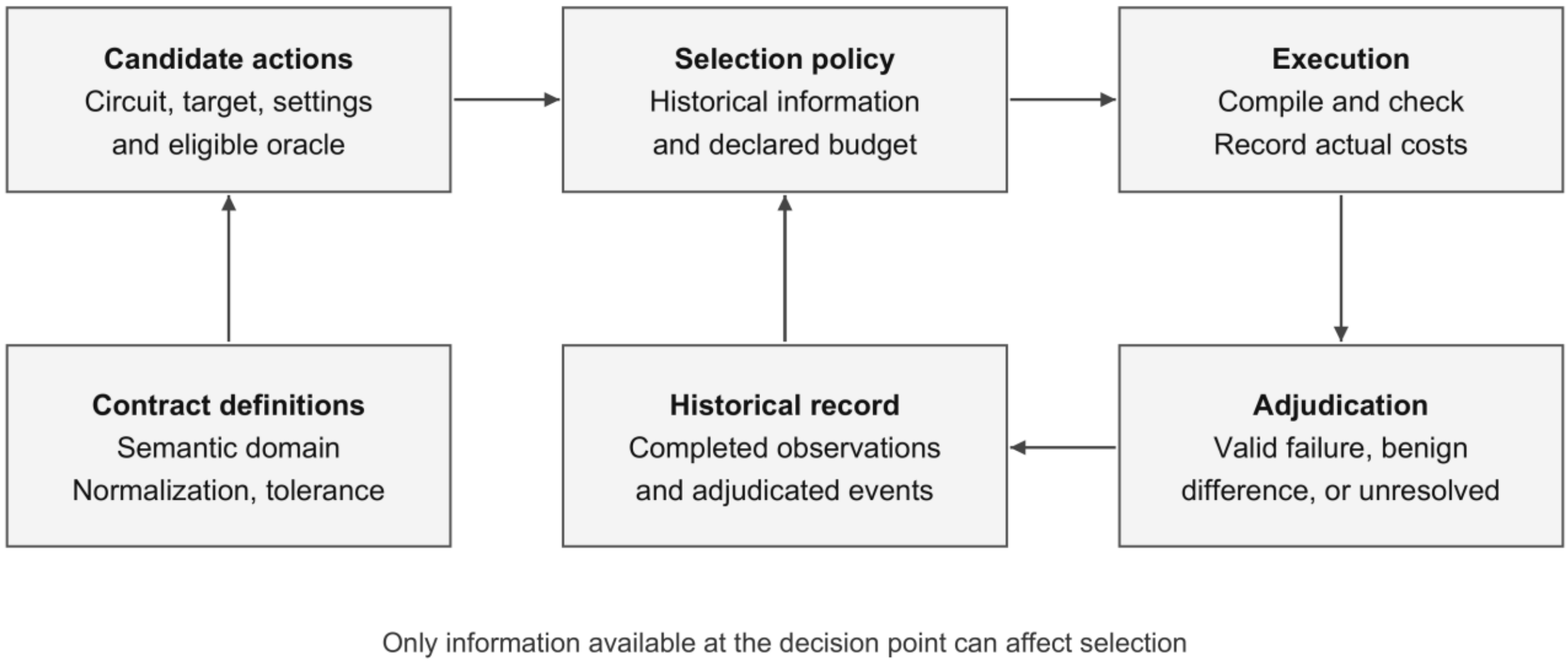

**Fig. 2** Proposed architecture for compiler regression testing. Contract definitions determine eligible observations and verdicts. Completed observations support later selections. The architecture itself has not been empirically evaluated in this review

### 7.4 Worked example of layout parameters and ancillary qubits

Consider a two-qubit source operation $U(\alpha, \beta) = R_x(\alpha) \otimes R_x(\beta)$, with the first tensor factor denoting logical qubit $q_0$. The compiler adds an ancilla initialized to zero and returns a physical circuit $V(\alpha, \beta)$, an input embedding E, and an output permutation record M. The contract requires the same logical operation for every permitted parameter binding and input, and requires the ancilla to be returned clean. For this example, the input placement is fixed and known independently. Output decoding $D_m$ is the permutation specified by the delivered metadata. Correctness means $D_m V(\alpha, \beta) E|\psi\rangle = U(\alpha, \beta)|\psi\rangle \otimes |0\rangle$, subject to the declared phase convention. This is a constructed example, not an executed compiler experiment.

Suppose V correctly places $q_0$ at physical output $p_1$ and $q_1$ at $p_0$, while M incorrectly claims the identity placement. A circuit checker using an independently supplied correct swap can accept V: it has answered whether the gate sequence is correct under that swap. The delivered pair (V, M) nevertheless violates the API contract. At $\alpha = \pi$, $\beta = 0$, and input $|00\rangle$, the ideal logical output is $|10\rangle$ up to phase. The physical values on $(p_0, p_1)$ are (0, 1); decoding with the incorrect identity record returns logical $|01\rangle$. This deterministic witness distinguishes the two interpretations. Equal bindings $\alpha = \beta$ would not distinguish this particular swap.

A checker that uses the delivered M to construct its semantic comparison can detect the same violation; metadata errors are not inherently invisible to every equivalence checker. The blind spot arises when the validation adapter substitutes a correct interpretation while the application receives the faulty one. An independent trace or specification can also validate M directly. Comparing only with last release's metadata is insufficient because a new correct routing may legitimately change it.

The same example separates four approaches. A verified pass [6, 7] supports its formalized transformation and prerequisites, but covers metadata serialization only if that operation belongs to its theorem. Layout-aware translation validation [8, 49] can check the delivered artifact when its adapter faithfully represents E and $D_m$. Executable testing [10, 13, 14] can expose the witness through labelled results, but passing selected bindings does not prove the parameter family. Drift monitoring [27] can flag a changed artifact or layout, but needs an independently justified contract before classifying the change as a defect. These are analytical applications of published mechanisms, not claims that the tools were run on this example.

Ancilla handling changes applicability. Equal dimensions of the full input and output operators cannot be assumed once an extra register is introduced. Here the comparison is restricted to ancilla input $|0\rangle$ and requires a clean output. If the actual contract permits discarding the ancilla, compare the reduced logical channel instead; that weaker obligation does not establish safe ancilla reuse. Likewise, a parameter cutoff applies only if the represented rotations and arithmetic satisfy its theorem [24]. A few arbitrary angle tests are not a substitute.

An operational decision procedure follows. First fix the input placement, delivered output interpretation, parameter domain, phase policy, ancilla policy, and target constraints. Next check input eligibility and successful compilation. Then choose a semantic method supporting those features and audit its conversions. Validate any delivered records outside that method's observation boundary separately. If the checker times out or lacks support, record an inconclusive semantic outcome; attempt another eligible representation or use declared-input tests such as the witness above. Such tests can establish a counterexample, but passing them supports only their restricted coverage. A required unresolved obligation prevents an overall assurance pass, even when all inexpensive checks succeed.

### 7.5 Applying the selection procedure to conditional gates and phase

The same five decisions used for the layout example apply to Cases A and D: specify the obligation; fix the relation and domain; establish applicability; select a distinguishing observation; and classify the evidence. The applications below are analytical reconstructions of failure mechanisms, not executions of the cited tools.

For a conditional merge, suppose the source applies $R_x(\alpha)$ when classical bit $c = 1$ and then applies $R_x(\beta)$ unconditionally. An incorrect candidate replaces both by unconditional $R_x(\alpha + \beta)$. The obligation is branch-wise preservation with the classical condition retained. At $c = 0$, $\alpha = \pi$, $\beta = 0$, and quantum input $|0\rangle$, the source yields $|0\rangle$ and the candidate yields $|1\rangle$ up to phase. Computational-basis measurement distinguishes them deterministically in the ideal model. A pass proof must include the conditional precondition; a dynamic equivalence checker must preserve the classical-quantum relation; an executable generator must include $c = 0$. Testing only the active branch would miss this witness. This construction illustrates Case A's applicability issue without asserting that it reproduces the exact Giallar artifact [7].

For phase-sensitive reuse, the obligation must first distinguish an isolated channel from an operation later placed under coherent control. $I$ and $-I$ satisfy the isolated-channel relation, so rejecting that replacement requires an additional API or composition requirement. Under coherent control, a control initialized in $|+\rangle$ becomes $|+\rangle$ for controlled-I and $|-\rangle$ for controlled-($-I$). A Hadamard followed by computational-basis measurement of the control yields 0 and 1 respectively. Exact operator comparison or this interference witness is appropriate to that contract; isolated output-distribution agreement is insufficient. This applies the reported phase issue [90] without inferring a repair from its closure.

In both cases, a counterexample establishes failure of the stated obligation for one artifact. Calling it a regression additionally requires a satisfying baseline under the same contract. The selection procedure thus determines both a suitable observation and the limit of the claim supported by that observation.

## 8 Evidence for validation across compiler revisions

The literature supports different parts of a regression workflow at different strengths. Giallar and compiler fuzzing provide compiler-fault evidence [7, 10, 11, 12, 13]. Circuit equivalence and property checking supply applicable mechanisms [3, 4, 8, 14]. Classical selection studies support transferred scheduling methodology [92, 93, 94, 95, 96, 97, 98, 99]. The drift workflow [27] directly monitors baseline-relative compilation changes, but its demonstrations do not establish fault-detection recall. The authors' preprint [34] studies repair records, not a prospectively evaluated selection policy. The extraction matrix records these categories separately; this survey establishes no common event corpus on which the complete architecture has been validated.

## 8.1 Distinguish three kinds of fault evidence

Table 9 distinguishes three evaluation objects. A forward regression event identifies a satisfying baseline and a violating candidate. A repair boundary demonstrates a defect and its correction, possibly without identifying its introduction. A synthetic mutation measures sensitivity to a controlled fault model. Each can inform evaluation, but their detection rates answer different questions.

**Table 9** Evidence categories for evaluation

| Category | Required provenance | Supported inference | Inference not supported by this evidence alone |
|---|---|---|---|
| Forward regression event | Satisfying baseline, violating candidate, contract, controlled configuration | Detection of deterioration at the observed change boundary | Population-wide prevalence from a small selected corpus |
| Bug-fix boundary | Failing parent, repairing change, reproducer, explanation | Detection of an existing defect and its repair | The original introducing event or prospective selection performance |
| Synthetic mutation | Parent artifact, mutation operator, validity and equivalence policy | Sensitivity to a controlled fault model | Real-world frequency or independence of related mutants |

For a forward event, both versions must be assessed under the same contract and a matched environment, unless environmental change is explicitly part of the event. Otherwise, the candidate failure may predate the change. Reversing a repair boundary can demonstrate an oracle's ability to distinguish versions, but does not reconstruct a naturally occurring introduction event or the information available at that time.

Fault adjudication should preserve uncertainty rather than hide it. A confirmed violation should have both an independently justified contract and a reproducible witness. An unresolved disagreement, unsupported feature, or unstable timing result should retain its own label instead of being folded into the same category. Developer confirmation can be useful evidence, but it may not be available and it cannot replace a clear specification. For that reason, an evaluation framework should publish its decision rules and the artifacts supporting them so that another evaluator can revisit the classification.

## 8.2 Information available at the selection point

Temporal leakage occurs when a selection policy uses information unavailable at its modeled decision time. Candidate source changes, circuit structure, backend snapshots, and prior outcomes can be legitimate inputs. Eventual failures, later repair patches, and retrospective root-cause labels cannot inform an earlier selection. Admissibility must therefore be specified for each feature rather than inferred from a dataset's name.

Evaluation partitions should group related fault variants and respect chronology. Mutants of one circuit, reproductions of one defect, and adjacent revisions of a repair can share predictive signals. Random row splitting can therefore overstate generalization. Feature normalization, threshold calibration, cost prediction, and hyperparameter selection must use only the designated training and validation partitions.

Adaptive selection may use an earlier test outcome once the observation has been obtained and its cost charged. Competing policies need comparable access to that feedback. Outcomes available only after subsequent execution remain inadmissible at the earlier decision point. Figure 3 separates these information boundaries.

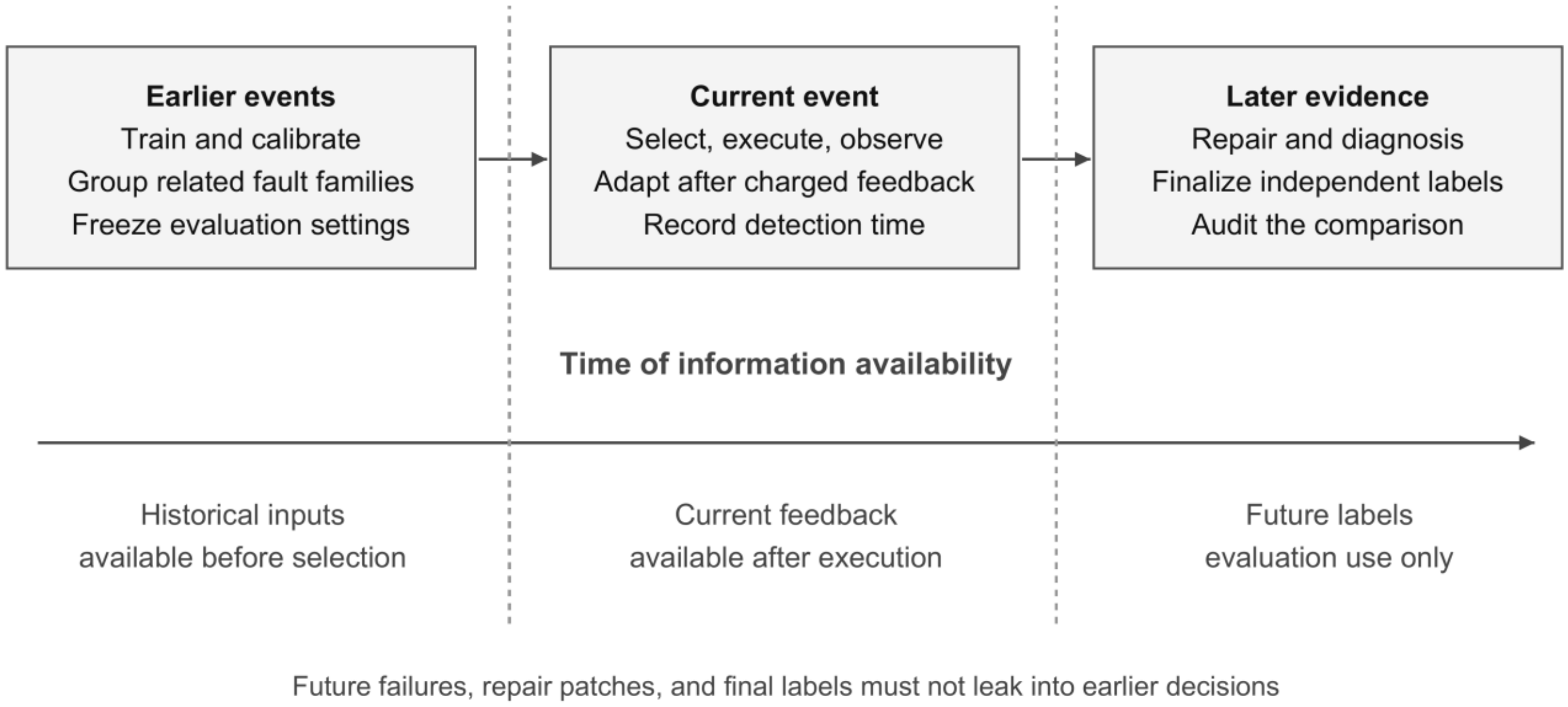

**Fig. 3** Information boundaries in an event-based evaluation. Earlier completed events inform training and calibration. Current observations become available only after their execution cost is incurred. Later repairs and final labels support evaluation but cannot influence earlier selection decisions

## 8.3 Comparisons and outcomes

A fair comparison fixes circuits, target snapshots, parameter domains, eligible oracles, and computational budgets. Random, cost-based, diversity-based, and historical policies provide interpretable baselines. Oracle additions and scheduling changes should be evaluated separately. Total cost includes compilation, conversion, simulation or shots, solving, and policy overhead [93, 101]. Supporting Information 4 details these evaluation requirements.

The primary quantity to track is valid detection of distinct compiler-change events under a stated budget, accompanied by time to first detection, false alarms, and resource use. Independent evidence should be counted at the level of the event family rather than the individual circuit variant. Measurement-based checks need an explicit error policy, and adaptive measurement must preserve its statistical guarantees. Timeouts and unsupported circuit features should remain separate from equivalence and inequivalence outcomes. With those controls in place, the evaluation can ask the useful question: whether additional observations of a quantum computation provide more assurance at an acceptable cost.

## 8.4 Verdict uncertainty and reliability claims

The review separates three sources of uncertainty because they require different treatments. Numerical tolerance belongs to the represented semantics and the arithmetic used to manipulate it. Sampling uncertainty comes from estimating quantities with finitely many observations. Implementation uncertainty arises from the compiler, checker, and adapters themselves. Exact inference removes shot noise, but only the second source is affected. The other two remain, just as additional shots cannot reveal a phase difference hidden by the chosen measurement basis [23, 39].

A sampled test should state its null hypothesis, practical discrepancy threshold, sample budget, and false-alarm policy before the experiment is run. Failure to reject an equality hypothesis is not, by itself, evidence of equivalence. If acceptance is defined by a distance falling below a tolerance, the procedure needs an appropriate bound or an equivalence-testing argument. Repeatedly inspecting a fixed-sample significance test and stopping when significance first appears can invalidate the nominal error rate. Adaptive stopping therefore requires a justified sequential rule. When many circuit comparisons are performed together, the campaign-level error criterion must also be stated. These are requirements for sound evaluation, not an assertion that every implementation reviewed here follows the same statistical procedure.

Fault discovery and operational reliability use different denominators. A fuzzer intentionally concentrates on unusual inputs, so the number of faults it discovers cannot be interpreted as the probability that a deployed compiler will fail on a user's normal workload. Estimating that probability requires an operational distribution over circuits, targets, and configurations, together with a valid failure criterion and a defensible sampling design. A simple analytical example makes the distinction concrete. With independent identically distributed trials and a perfect failure oracle, observing zero failures in n trials gives the exact one-sided 95 percent binomial upper bound $1 - 0.05^{(1/n)}$ on the trial failure probability, obtained by setting $(1 - p)^n = 0.05$. That result does not carry over unchanged when tests are adaptively generated or dependent, environments drift, or the oracle has an unknown miss rate.

For the studies included in this review, the defensible outcomes are scoped defect detections, supported verification claims, and measured resource use. They do not provide the operational evidence needed for a reliability estimate. A regression report should therefore list eligible comparisons, adjudicated failures, inconclusive outcomes, and excluded inputs separately. Keeping those denominators visible prevents a lower apparent failure fraction from being produced simply by removing difficult cases from the count.

### 8.5 Illustrative attribution across compiler revisions

Extend Section 7.4 with hypothetical revisions $r_0$ and $r_1$. Both produce the same correctly swapped physical circuit; $r_0$ returns the swap record and $r_1$ returns the identity record. Fix the source, parameter binding $(\pi, 0)$, input $|00\rangle$, target snapshot, seed, dependencies, and checker implementation. The declared obligation is preservation of labelled logical outputs using the delivered metadata. Retain the source and output artifacts, mappings, environment identifiers, statuses, and oracle interpretation for both revisions.

Algebraically, decoding $r_0$ yields $|10\rangle$ and decoding $r_1$ yields $|01\rangle$. Thus $r_0$ satisfies this test obligation and $r_1$ violates it. A gate comparison that substitutes the correct swap for both revisions passes both and cannot supply this verdict. A result-label check using each delivered mapping fails $r_1$. This demonstrates the role of an additional observation without inventing a measured detection rate. It also does not prove all-input correctness of $r_0$ from one passing test.

If those artifacts were observed in an actual matched experiment, they would establish deterioration between the two revisions for the stated obligation. To identify an introducing commit within a longer interval, eligible intermediate revisions would need checking and the candidate change would require adjudication; bisection alone is unsafe when failures are non-monotonic. If dependencies or backend descriptions also change, attribute the result to the combined toolchain change until those factors are controlled. If $r_0$ also fails, the record is an existing defect, not a newly demonstrated regression. If either comparison is inconclusive, attribution remains unresolved.

This scenario explains the architecture's proposed operation. It supplies no real release identifiers, runtime measurements, or claim that an actual compiler history was reproduced. The evaluation requirements in Sections 8.2–8.4 are intended to make a future implementation testable.

## 9 Research agenda and threats to validity

The research agenda follows the five findings in Section 6.5. Existing methods already address parameterized and dynamic semantics. The remaining integration questions concern the guarantees that survive actual compiler interfaces, the inputs needed to exercise applicability boundaries, and the distinction between processing scale and semantic assurance.

### 9.1 Open problems in quantum compiler assurance

The synthesis identifies four connected evaluation needs. Compositional validation must carry phase, layout, initialization, and ancilla assumptions across intermediate representations. Dynamic-circuit checks must preserve the association between classical records and conditional quantum behavior. Parameter-domain and approximation claims need explicit supported syntax, arithmetic, and composition rules. Scalable portfolios must report semantic coverage separately from target compliance and compilation performance. Existing methods address parts of these questions; the unresolved issue is reliable integration under actual compiler interfaces.

**Table 10** Research questions and required evidence

| Research question | Required evidence | Boundary of a positive result |
|---|---|---|
| Does correctness survive control or ancilla reuse? | Explicit composition contracts and distinguishing witnesses | An isolated-circuit result covers only its allowed contexts |
| Are classical records preserved? | Labelled conditional-process comparisons | Averaged channels can conceal record mismatches |
| Does validation cover the parameter domain? | Applicable theorem or explicitly restricted binding tests | Arbitrary sampled agreement is not universal equality |
| Are approximation budgets compositional? | A fixed metric and justified accumulation rule | Different metrics cannot be pooled without a bound |
| Do interface observations add value? | Unique adjudicated detections, false alarms, costs, and ablations | Stronger acceptance can also overconstrain valid outputs |
| Does the portfolio transfer across revisions? | Independent change events and held-out settings | Related variants of one defect do not establish transfer |

## 9.2 Oracle complementarity and transfer

The main empirical question is the value of an additional observation under a fixed budget. Evaluate the same eligible events with and without that observation, preserving the scheduler where possible, and report unique detections, overlap, false alarms, and unresolved outcomes. Changes to scheduling should be evaluated separately from changes to the oracle set. Historical cost or risk models also need chronological validation: a new setting of a known fault is weaker transfer evidence than a new fault mechanism.

Computational reuse creates a further trade-off. Shared compilation artifacts and intermediate states reduce cost but may couple checker failures. A shared faulty adapter can defeat several nominally different methods. Future evaluations should record these dependencies and compare simpler cost- or diversity-based policies before attributing gains to more elaborate scheduling.

## 9.3 Threats to validity

Selection and accessibility limit the reach of a critical integrative review. Topic searches and selective reference following naturally favour established terminology, highly cited work, and material that is easy to obtain. The seven-study pilot and its narrow compiler-testing scope can also shape the direction of later searches. Historical aggregate counts lack record-level support, so they cannot establish coverage or screening reliability. Searching for adjacent mechanisms and consulting recent reviews reduces this bias but cannot remove it. Exhaustive coverage was not the objective. The synthesis therefore supports analytical relationships among the methods included in the review, and no broader prevalence claim. It does not estimate how common any approach is, nor can it demonstrate that every alternative technique is absent.

Extraction introduces a separate source of interpretive uncertainty. Different publication versions, artifact descriptions, and terminology may describe the same underlying method at different levels of scope. We limited this problem by linking publication families through persistent identifiers, separating formal guarantees from experimental findings, and recording whether a claim was checked against full text or only an abstract-level record. Screening and coding were not independently duplicated, and the reported experiments were not rerun. These limitations are stated directly. Ambiguous guarantees were therefore interpreted conservatively, and claims based on limited access are restricted to the description actually inspected. The procedure improves traceability, but independent extraction or replication can still correct the resulting interpretation and remains valuable.

Cross-study inference is constrained by substantial heterogeneity. Circuit families, compiler versions, numerical policies, resource settings, and fault definitions vary across the reviewed work. Pooling runtimes or bug counts across studies with these differences would confound the settings rather than average them meaningfully. The synthesis instead compares obligations and evidence types, keeping each numerical result attached to the evaluation that produced it. Independence is also reduced by shared tool lineages and related benchmark instances. Repeating the same study is not independent evidence, and multiple variants of one defect do not become independent observations of fault prevalence or transfer simply because they are numerous.

The method-selection guidance and regression architecture are analytical contributions. This distinction matters: the review does not implement a complete contract-checking system, measure its cost, or validate the architecture on independent compiler-change events. Those tasks remain future work. The conclusions drawn from compilation and simulation also transfer only indirectly to noisy device execution, where calibration and measurement introduce additional confounders. Emerging preprints are included because they can expose relevant mechanisms before independent replication is available. Their inclusion broadens coverage. It is not evidence of maturity and does not strengthen the underlying evidence by itself.

## 10 Conclusion

Quantum-compiler quality requires evidence that matches the operational requirement being assessed. Semantic preservation, executable validity, output interpretation, and performance objectives interact, but none can substitute for another. A verification result has meaning only together with its input quantification, program domain, numerical assumptions, and interface boundary. An executable test likewise depends on its observations, reference relation, and uncertainty. If those conditions are omitted, neither type of result supports a well-defined claim about the compiler.

The formal results, documented failures, and analytical example distinguish limitations addressed by semantic reasoning, input generation, and interface observations; they do not quantify the benefit of a combined portfolio. The conditional-gate and parameter-order defects show why applicability and input discrimination matter. The shared-crash and phase-reuse cases show why combined checks need an explicit contract. Published defect totals, mutation experiments, and large-circuit benchmarks also represent different forms of evidence. Treating them as inputs to one effectiveness ranking would change the meaning of what the original evaluations measured.

For assurance across compiler revisions, the synthesis proposes three practical steps. Combine checks only when their requirements are compatible. Preserve the context needed to interpret every verdict. Evaluate each check by what it contributes on independently attributed change events. The method-selection guide makes these choices explicit, but it is not presented as a validated optimal policy. This review has not established one. Further progress requires transparent method comparisons, reliable interpretation of compiler interfaces, and evidence that additional checking effort finds distinct, valid failures within the available resource budget.

## Statements and Declarations

### Declaration of Generative AI and AI-Assisted Technologies in the Manuscript Preparation Process

During preparation of this work, the authors used a generative AI assistant for language editing, prose polishing, and consistency checking across manuscript drafts, and to generate the plotting code for Figures 1-3. These figures are conceptual diagrams and analytical illustrations rather than plots of new empirical measurements. After using this tool, the authors reviewed and edited the content as needed and take full responsibility for the content of the published article. Generative AI was not used to fabricate, alter, or select any data or empirical results reported in this paper.

Funding: This research received no specific grant from any funding agency in the public, commercial, or not-for-profit sectors.

Competing interests: The authors declare no competing interests.

Author contributions: This work forms part of Furqan Nasir’s PhD research in Computer Science at City University of Science and Information Technology (CUSIT), Peshawar, Pakistan, under the supervision of Muhammad Arif Shah and the co-supervision of Iftikhar Alam.

Data availability: The search archive, candidate decisions, catalogue selection reasons, evidence classifications, finding ledger, and documented case records accompany this review. The review protocol, historical SLR provenance and seed mapping, and search updates are supplied in Supporting Information 1; the source catalogue and focused extraction are supplied in Supporting Information 2; documented case records are supplied in Supporting Information 3. No new quantum compiler benchmark measurements were generated.

## Supplementary information

Supporting Information 1: Review protocol, historical SLR provenance and seed mapping, search queries, archived search returns, and candidate decisions

Supporting Information 2: Source catalogue, selection decisions, technical extraction, and finding ledger

Supporting Information 3: Documented cases and boundaries of their interpretation

Supporting Information 4: Coverage update, worked-example derivation, and evaluation requirements

## References

[1] F. Javier Cardama, Jorge Vázquez-Pérez, Tomás F. Pena, Juan C. Pichel, Andrés Gómez. Quantum Compilation Process: A Survey. *Euro-Par 2024: Parallel Processing Workshops. Lecture Notes in Computer Science*. 2025; 15385; 100–112. https://doi.org/10.1007/978-3-031-90200-0_9

[2] F. Javier Cardama, Jorge Vázquez-Pérez, César Piñeiro, Juan C. Pichel, Tomás F. Pena, Andrés Gómez. Review of intermediate representations for quantum computing. *The Journal of Supercomputing*. 2025; 81; 418. https://doi.org/10.1007/s11227-024-06892-2

[3] Tian-Fu Chen, Jie-Hong R. Jiang, Min-Hsiu Hsieh. Partial Equivalence Checking of Quantum Circuits. *2022 IEEE International Conference on Quantum Computing and Engineering (QCE)*. 2022; 594–604. https://doi.org/10.1109/QCE53715.2022.00082

[4] Tom Peham, Lukas Burgholzer, Robert Wille. Equivalence Checking of Parameterized Quantum Circuits. *Proceedings of the 28th Asia and South Pacific Design Automation Conference*. 2023; 702-708. https://doi.org/10.1145/3566097.3567932

[5] Liam Hurwitz, Kamalika Datta, Abhoy Kole, Rolf Drechsler. Is Simulation the only Alternative for Effective Verification of Dynamic Quantum Circuits? *Lecture Notes in Computer Science*. 2024; 201-217. https://doi.org/10.1007/978-3-031-62076-8_13

[6] Kesha Hietala, Robert Rand, Shih-Han Hung, Xiaodi Wu, Michael Hicks. A Verified Optimizer for Quantum Circuits. *Proceedings of the ACM on Programming Languages*. 2021; 5; POPL; 1-29. https://doi.org/10.1145/3434318

[7] Runzhou Tao, Yunong Shi, Jianan Yao, Xupeng Li, Ali Javadi-Abhari, Andrew W. Cross, Frederic T. Chong, Ronghui Gu. Giallar: Push-button Verification for the Qiskit Quantum Compiler. *Proceedings of the 43rd ACM SIGPLAN International Conference on Programming Language Design and Implementation*. 2022; 641-656. https://doi.org/10.1145/3519939.3523431

[8] Lukas Burgholzer, Robert Wille. Advanced Equivalence Checking for Quantum Circuits. *IEEE Transactions on Computer-Aided Design of Integrated Circuits and Systems*. 2021; 40; 9; 1810-1824. https://doi.org/10.1109/tcad.2020.3032630

[9] Tom Peham, Lukas Burgholzer, Robert Wille. Equivalence Checking of Quantum Circuits with the ZX-Calculus. *IEEE Journal on Emerging and Selected Topics in Circuits and Systems*. 2022. https://doi.org/10.1109/JETCAS.2022.3202204

[10] Jiyuan Wang, Qian Zhang, Guoqing Harry Xu, Miryung Kim. QDiff: Differential Testing of Quantum Software Stacks. *2021 36th IEEE/ACM International Conference on Automated Software Engineering (ASE)*. 2021; 692-704. https://doi.org/10.1109/ase51524.2021.9678792

[11] Matteo Paltenghi, Michael Pradel. MorphQ: Metamorphic Testing of the Qiskit Quantum Computing Platform. *2023 IEEE/ACM 45th International Conference on Software Engineering (ICSE)*. 2023; 2413-2424. https://doi.org/10.1109/icse48619.2023.00202

[12] Junjie Luo, Shangzhou Xia, Fuyuan Zhang, Jianjun Zhao. QEMI: A Quantum Software Stacks Testing Framework via Equivalence Modulo Inputs. *Fundamental Approaches to Software Engineering. FASE 2026. Lecture Notes in Computer Science*. 2026; 16504; 149–169. https://doi.org/10.1007/978-3-032-22774-4_8

[13] Ilan Iwumbwe, Benny Zong Liu, John Wickerson. QuteFuzz: Fuzzing quantum compilers using randomly generated circuits with control flow and subcircuits. *Programming Languages for Quantum Computing Workshop (PLanQC 2025)*. 2025. https://johnwickerson.github.io/papers/QuteFuzz.pdf

[14] Gabriel Pontolillo, Mohammad Reza Mousavi, Marek Grzesiuk. QuCheck: A Property-based Testing Framework for Quantum Programs in Qiskit. *ACM Transactions on Quantum Computing*. 2026. Advance online publication, 8 May 2026. https://doi.org/10.1145/3815169 Earlier mechanism and evaluation version inspected: https://arxiv.org/abs/2503.22641v1

[15] Paul D. Nation, Abdullah Ash Saki, Sebastian Brandhofer, Luciano Bello, Shelly Garion, Matthew Treinish, Ali Javadi-Abhari. Benchmarking the performance of quantum computing software for quantum circuit creation, manipulation and compilation. *Nature Computational Science*. 2025; 5; 5; 427-435. https://doi.org/10.1038/s43588-025-00792-y

[16] Huynh Khanh Vi Tran, Nauman bin Ali, Michael Unterkalmsteiner, Jürgen Börstler, Panagiota Chatzipetrou. Quality attributes of test cases and test suites – importance & challenges from practitioners' perspectives. *Software Quality Journal*. 2025; 33; 1; 9. https://doi.org/10.1007/s11219-024-09698-w

[17] Earl T. Barr, Mark Harman, Phil McMinn, Muzammil Shahbaz, Shin Yoo. The Oracle Problem in Software Testing: A Survey. *IEEE Transactions on Software Engineering*. 2015; 41; 5; 507-525. https://doi.org/10.1109/tse.2014.2372785

[18] Matteo Paltenghi; Michael Pradel. A Survey on Testing and Analysis of Quantum Software. Preprint. 2024. https://arxiv.org/abs/2410.00650v1 Accessed 19 September 2026.

[19] Neilson Carlos Leite Ramalho, Higor Amario de Souza, Marcos Lordello Chaim. Testing and Debugging Quantum Programs: The Road to 2030. *ACM Transactions on Software Engineering and Methodology*. 2025; 34; 5; 1-46. https://doi.org/10.1145/3715106

[20] Evandro Rosa, Rafael Santiago. Schrödinger's bug: a survey on quantum software debugging. *Quantum Information Processing*. 2026; 25; 3; 96. https://doi.org/10.1007/s11128-026-05114-3

[21] Yuechen Li, Minqi Shao, Jianjun Zhao, Qichen Wang. A Methodological Analysis of Empirical Studies in Quantum Software Testing. *ACM Transactions on Software Engineering and Methodology*. 2026. https://doi.org/10.1145/3819590 Review structure inspected in https://arxiv.org/abs/2601.08367v2

[22] Xinyi Wang, Shaukat Ali, Davide Taibi. The Landscape of Quantum Software Testing Tools. *IEEE Software*. 2025; 42; 5; 136-140. https://doi.org/10.1109/ms.2025.3578154

[23] Jiaming Ye, Xiongfei Wu, Shangzhou Xia, Fuyuan Zhang, Jianjun Zhao. Is Measurement Enough? Rethinking Output Validation in Quantum Program Testing. *2025 40th IEEE/ACM International Conference on Automated Software Engineering (ASE)*. 2025; 3824-3828. https://doi.org/10.1109/ase63991.2025.00324

[24] Neil J. Ross; Scott Wesley. Cutoff Theorems for the Equivalence of Parameterized Quantum Circuits. LIPIcs MFCS. 2025. 345:85:1–85:19. https://doi.org/10.4230/LIPIcs.MFCS.2025.85

[25] Amanda Xu; Abtin Molavi; Lauren Pick; Swamit Tannu; Aws Albarghouthi. Synthesizing Quantum-Circuit Optimizers. Proceedings of the ACM on Programming Languages. 2023. 7(PLDI):835–859. https://doi.org/10.1145/3591254

[26] Nengkun Yu; Xuan Du Trinh; Thomas Reps. Scalable Equivalence Checking and Verification of Shallow Quantum Circuits. Proceedings of the ACM on Programming Languages. 2025. 9(OOPSLA2):2870–2897. https://doi.org/10.1145/3763153

[27] Jouni Peltonen; Otso Kinanen; Vlad Stirbu. Catching Transpilation Drift with a CI/CD Workflow in Quantum Software Development. Preprint. 2026. Accessed 19 September 2026. https://arxiv.org/abs/2608.08248v1

[28] Ziming Zhao; Tingting Li; Zhaoxuan Li; Jianwei Yin. Relational Verification for Cost-Aware Quantum Program Optimization. *Proceedings of the AAAI Conference on Artificial Intelligence*. 2026; 40(17):14414–14422. https://doi.org/10.1609/aaai.v40i17.38457

[29] Yu-Fang Chen; Kai-Min Chung; Ondřej Lengál; Jyun-Ao Lin; Wei-Lun Tsai. AutoQ: An Automata-Based Quantum Circuit Verifier. *Computer Aided Verification*. 2023; LNCS 13966. https://doi.org/10.1007/978-3-031-37709-9_7

[30] Daisuke Sakamoto; Soshun Naito; Yusei Mori; Kosuke Mitarai. Efficient equivalence checking of Clifford-U circuits with shared single-qubit unitaries. arXiv preprint arXiv:2603.12697. 2026. https://arxiv.org/abs/2603.12697

[31] Chao Lu; Navnil Choudhury; Utsav Banerjee; Abdullah Ash Saki; Kanad Basu. QuBEC: Boosting Equivalence Checking for Quantum Circuits with QEC Embedding. *IEEE Transactions on Computer-Aided Design of Integrated Circuits and Systems*. 2024; 43(7):2037–2042. https://doi.org/10.1109/TCAD.2024.3361402

[32] Lukas Burgholzer; Robert Wille. Improved DD-based Equivalence Checking of Quantum Circuits. *Asia and South Pacific Design Automation Conference*. 2020. https://research.jku.at/en/publications/improved-dd-based-equivalence-checking-of-quantum-circuits/

[33] Kesha Hietala; Robert Rand; Liyi Li; Shih-Han Hung; Xiaodi Wu; Michael Hicks. A Verified Optimizer for Quantum Circuits. ACM Transactions on Programming Languages and Systems. 2023. 45(3):18, 35 pages. https://doi.org/10.1145/3604630

[34] Furqan Nasir; Arif Shah; Iftikhar Alam. What Output-Equivalence Oracles Miss: An Empirical Study of Equivalence-Invisible Bug Fixes in Quantum Transpilers (Qiskit, tket, Cirq). Preprint. 2026. Accessed 19 September 2026. https://arxiv.org/abs/2609.13839v1

[35] Antonio García de la Barrera; Ignacio García-Rodríguez de Guzmán; Macario Polo; Mario Piattini. Quantum software testing: State of the art. Journal of Software: Evolution and Process. 2023. 35(4):e2419. First online 2021. https://doi.org/10.1002/smr.2419

[36] Arun Govindankutty. A Review of Formal Methods in Quantum-Circuit Verification. Electronics. 2026. 15(5):1125. https://doi.org/10.3390/electronics15051125

[37] Emma Andrews, Aruna Jayasena, Prabhat Mishra. A Survey of Functional Testing and Validation of Quantum Circuits. *IEEE Design & Test*. 2026; 43(5):5–19. Online record available; October 2026 issue forthcoming at the cutoff. https://doi.org/10.1109/mdat.2026.3698368

[38] Rodolfo Gil-Pereira; Ronnie de Souza Santos; Cleyton Magalhaes; Italo Santos. Exploring Quantum Software Testing Across Research and Practice: Emerging Results from a Multivocal Literature Review. Preprint. 2026. https://arxiv.org/abs/2609.00354v1 Accessed 19 September 2026.

[39] Kanguk Lee, Jaemin Hong, Sukyoung Ryu. Exact Inference for Quantum Circuits: A Testing Oracle for Quantum Software Stacks. *2025 40th IEEE/ACM International Conference on Automated Software Engineering (ASE)*. 2025; 2465-2477. https://doi.org/10.1109/ase63991.2025.00203

[40] Noah H. Oldfield, Christoph Laaber, Tao Yue, Shaukat Ali. Faster and Better Quantum Software Testing through Specification Reduction and Projective Measurements. *ACM Transactions on Software Engineering and Methodology*. 2025; 34; 7; 1-39. https://doi.org/10.1145/3714468

[41] Jérôme Ricciardi; Sébastien Bardin; Christophe Chareton; Benoît Valiron. Quantum Circuit Equivalence Checking: A Tractable Bridge From Unitary to Hybrid Circuits. Preprint. 2025. https://arxiv.org/abs/2511.22523v1 Accessed 19 September 2026.

[42] Xin Hong; Mingsheng Ying; Yuan Feng; Xiangzhen Zhou; Sanjiang Li. Approximate Equivalence Checking of Noisy Quantum Circuits. Preprint. 2021. https://arxiv.org/abs/2103.11595v2 Accessed 19 September 2026.

[43] Mingyu Huang; Ji Guan; Wang Fang; Mingsheng Ying. Approximation Methods for Simulation and Equivalence Checking of Noisy Quantum Circuits. Preprint. 2025. https://arxiv.org/abs/2503.10340v2 Accessed 19 September 2026.

[44] Yunong Shi; Runzhou Tao; Xupeng Li; Ali Javadi-Abhari; Andrew W. Cross; Frederic T. Chong; Ronghui Gu. CertiQ: Mostly-automated Verification of a Realistic Quantum Compiler. Preprint. 2020. https://arxiv.org/abs/1908.08963v5 Accessed 19 September 2026.

[45] Liyi Li; Finn Voichick; Kesha Hietala; Yuxiang Peng; Xiaodi Wu; Michael Hicks. Verified Compilation of Quantum Oracles. Proceedings of the ACM on Programming Languages. 2022; 6(OOPSLA2):146. https://doi.org/10.1145/3563309

[46] Liyi Li; Fenfen An; Federico Zahariev; Zhi Xiang Chong; Amr Sabry; Mark S. Gordon. A Verified Compiler for Quantum Simulation. arXiv preprint arXiv:2509.18583v3. 2025. https://arxiv.org/abs/2509.18583v3

[47] Matthew Amy; Martin Roetteler; Krysta M. Svore. Verified Compilation of Space-Efficient Reversible Circuits. Computer Aided Verification. 2017. https://doi.org/10.1007/978-3-319-63390-9_1

[48] Robert Rand; Jennifer Paykin; Dong-Ho Lee; Steve Zdancewic. ReQWIRE: Reasoning about Reversible Quantum Circuits. Electronic Proceedings in Theoretical Computer Science. 2019; 287:299-312. https://doi.org/10.4204/EPTCS.287.17

[49] Lukas Burgholzer, Rudy Raymond, Robert Wille. Verifying Results of the IBM Qiskit Quantum Circuit Compilation Flow. *2020 IEEE International Conference on Quantum Computing and Engineering (QCE)*. 2020; 356-365. https://doi.org/10.1109/qce49297.2020.00051

[50] Hsiao-Lun Liu, Yi-Ting Li, Yung-Chih Chen, Chun-Yao Wang. A Robust Approach to Detecting Non-Equivalent Quantum Circuits Using Specially Designed Stimuli. *Proceedings of the 28th Asia and South Pacific Design Automation Conference*. 2023; 696-701. https://doi.org/10.1145/3566097.3567935

[51] Chun-Yu Wei, Yuan-Hung Tsai, Chiao-Shan Jhang, Jie-Hong R. Jiang. Accurate BDD-based unitary operator manipulation for scalable and robust quantum circuit verification. *Proceedings of the 59th ACM/IEEE Design Automation Conference*. 2022; 523-528. https://doi.org/10.1145/3489517.3530481

[52] Sebastiaan Brand; Alfons Laarman. Q-Sylvan: A Parallel Decision Diagram Package for Quantum Computing. Preprint. 2025. https://arxiv.org/abs/2508.00514v1 Accessed 19 September 2026.

[53] Jingyi Mei, Tim Coopmans, Marcello Bonsangue, Alfons Laarman. Equivalence Checking of Quantum Circuits by Model Counting. *Lecture Notes in Computer Science*. 2024; 401-421. https://doi.org/10.1007/978-3-031-63501-4_21

[54] Wei-Jia Huang, Christophe Chareton, Yu-Fang Chen, Kai-Min Chung, Min-Hsiu Hsieh, Alfons Laarman, Jingyi Mei. Equivalence Checking of Quantum Circuits via Path-Sum and Weighted Model Counting. *Lecture Notes in Computer Science*. 2026; 419-439. https://doi.org/10.1007/978-3-032-22749-2_21

[55] Xuan Du Trinh; Meghana Sistla; Nengkun Yu; Thomas Reps. Manjushri: A Tool for Equivalence Checking of Quantum Circuits. Preprint. 2026. https://arxiv.org/abs/2601.22372v1 Accessed 19 September 2026.

[56] Chao Lu; Navnil Choudhury; Kanad Basu. QuaSi: A Scalable and Reliable Quantum Simulation-based Equivalence Checking Framework. IEEE QCE 2024. https://doi.org/10.1109/QCE60285.2024.00123

[57] Karol Bartkiewicz; Patrycja Tulewicz. Context-Verified, Error-Budget-Aware Decomposition Selection for Toffoli Networks. arXiv preprint arXiv:2606.31791v1. 2026. https://arxiv.org/abs/2606.31791v1

[58] Xin Hong; Wei-Jia Huang; Wei-Chen Chien; Yuan Feng; Min-Hsiu Hsieh; Sanjiang Li; Mingsheng Ying. Equivalence Checking of Parameterised Quantum Circuits. Preprint. 2024. https://arxiv.org/abs/2404.18456v1 Accessed 19 September 2026.

[59] Lukas Burgholzer; Robert Wille. Handling Non-Unitaries in Quantum Circuit Equivalence Checking. *Proceedings of the 59th ACM/IEEE Design Automation Conference*. 2022. https://doi.org/10.1145/3489517.3530482

[60] Yu-Fang Chen, Kai-Min Chung, Min-Hsiu Hsieh, Wei-Jia Huang, Ondřej Lengál, Jyun-Ao Lin, Wei-Lun Tsai. AutoQ 2.0: From Verification of Quantum Circuits to Verification of Quantum Programs. *Tools and Algorithms for the Construction and Analysis of Systems*. 2025; 15698; 87–108. https://doi.org/10.1007/978-3-031-90660-2_5

[61] Parosh Aziz Abdulla, Yu-Fang Chen, Michal Hečko, Lukáš Holík, Ondřej Lengál, Jyun-Ao Lin, Ramanathan S. Thinniyam. Parameterized Verification of Quantum Circuits. *Proceedings of the ACM on Programming Languages*. 2026; 10(POPL):70:1–70:30. https://doi.org/10.1145/3776712

[62] Ziyuan Wang; Bin Cheng; Longxiang Yuan; Zhengfeng Ji. FeynmanDD: Quantum Circuit Analysis with Classical Decision Diagrams. Computer Aided Verification. 2025. LNCS 15934:28–52. https://doi.org/10.1007/978-3-031-98685-7_2

[63] Aaron Sander; Lukas Burgholzer; Robert Wille. Equivalence checking of quantum circuits via intermediary matrix product operator. Physical Review Research. 2025. 7:023261. https://doi.org/10.1103/3q71-y8cf

[64] Xin Hong; Yuan Feng; Sanjiang Li; Mingsheng Ying. Equivalence Checking of Dynamic Quantum Circuits. Preprint. 2021. Author version; accessed 19 September 2026. https://arxiv.org/abs/2106.01658v1

[65] Liyi Li; Mingwei Zhu; Rance Cleaveland; Alexander Nicolellis; Yi Lee; Le Chang; Xiaodi Wu. Qafny: A Quantum-Program Verifier. LIPIcs ECOOP. 2024. 313:24:1–24:31. https://doi.org/10.4230/LIPIcs.ECOOP.2024.24

[66] Feifei Cheng; Sushen Vangeepuram; Henry Allard; Seyed Mohammad Reza Jafari; Alex Potanin; Liyi Li. Embedding Quantum Program Verification into Dafny. *Proceedings of the ACM on Programming Languages. 9(OOPSLA2):379*. 2025. https://doi.org/10.1145/3763157

[67] Suhaib Al-Rousan, Kim Guldstrand Larsen, Christian Schilling, Max Tschaikowski. Reinforcement Learning for Efficient Quantum Circuit Equivalence Checking via Tensor Network Contraction. *AI in Science Summit 2025*. 2025. https://vbn.aau.dk/en/publications/reinforcement-learning-for-efficient-quantum-circuit-equivalence-/

[68] Navnil Choudhury, Ameya S. Bhave, Kanad Basu. ZXNet: ZX Calculus-Driven Graph Neural Network Framework for Quantum Circuit Equivalence Checking. *62nd ACM/IEEE Design Automation Conference (DAC)*. 2025. https://62dac.conference-program.com/presentation/?id=RESEARCH452&sess=sess125

[69] Linsey Kitt, Myra B. Cohen. MorphQ++: A Reproducibility Study of Metamorphic Testing on Quantum Compilers. *Proceedings of the 2024 Workshop on Replications and Negative Results*. 2024; 15-21. https://doi.org/10.1145/3695750.3695823

[70] Jiaming Ye; Fuyuan Zhang; Shangzhou Xia; Xiaoyu Guo; Xiongfei Wu; Jianjun Zhao; Yinxing Xue. QSPE: Enumerating Skeletal Quantum Programs for Quantum Library Testing. Preprint. 2026. https://arxiv.org/abs/2602.00024v1 Accessed 19 September 2026.

[71] Matteo Paltenghi, Michael Pradel. IterTestQ: Assembly-Level, Cross-Platform Testing of Quantum Computing Platforms. *Proceedings of the ACM on Software Engineering*. 2026; 3; ISSTA; Article ISSTA120, 23 pages. Forthcoming; October 2026 issue date. https://doi.org/10.1145/3832211

[72] Chunqiu Steven Xia; Matteo Paltenghi; Jia Le Tian; Michael Pradel; Lingming Zhang. Fuzz4All: Universal Fuzzing with Large Language Models. ICSE 2024. https://doi.org/10.1145/3597503.3639121

[73] Chan Gu Kang, Joonghoon Lee, Hakjoo Oh. Statistical Testing of Quantum Programs via Fixed-Point Amplitude Amplification. *Proceedings of the ACM on Programming Languages*. 2024; 8; OOPSLA2; 140-164. https://doi.org/10.1145/3689716

[74] Peixun Long; Jianjun Zhao. Equivalence, identity, and unitarity checking in black-box testing of quantum programs. *Journal of Systems and Software. 211:112000*. 2024. https://doi.org/10.1016/j.jss.2024.112000

[75] Xinyi Wang; Paolo Arcaini; Tao Yue; Shaukat Ali. QuCAT: A Combinatorial Testing Tool for Quantum Software. *Proceedings of the 38th IEEE/ACM International Conference on Automated Software Engineering*. 2023. https://doi.org/10.1109/ASE56229.2023.00062

[76] Peixun Long; Jianjun Zhao. Testing Multi-Subroutine Quantum Programs: From Unit Testing to Integration Testing. *ACM Transactions on Software Engineering and Methodology. 33(6)*. 2024. https://doi.org/10.1145/3656339

[77] Navnil Choudhury, Ameya Bhave, Kanad Basu. Concolic Testing for Quantum Compilers. *2025 IEEE 43rd International Conference on Computer Design (ICCD)*. 2025; 348-355. https://doi.org/10.1109/iccd65941.2025.00057

[78] Matteo Paltenghi, Michael Pradel. Analyzing Quantum Programs with LintQ: A Static Analysis Framework for Qiskit. *Proceedings of the ACM on Software Engineering*. 2024; 1; FSE; 2144-2166. https://doi.org/10.1145/3660802

[79] Gabriel Joseph Pontolillo, Mohammad Reza Mousavi. Delta Debugging for Property-Based Regression Testing of Quantum Programs. *Proceedings of the 5th ACM/IEEE International Workshop on Quantum Software Engineering*. 2024; 1-8. https://doi.org/10.1145/3643667.3648219

[80] Enaut Mendiluze, Shaukat Ali, Paolo Arcaini, Tao Yue. Muskit: A Mutation Analysis Tool for Quantum Software Testing. *2021 36th IEEE/ACM International Conference on Automated Software Engineering (ASE)*. 2021; 1266-1270. https://doi.org/10.1109/ase51524.2021.9678563

[81] Eñaut Mendiluze Usandizaga, Shaukat Ali, Tao Yue, Paolo Arcaini. Quantum circuit mutants: Empirical analysis and recommendations. *Empirical Software Engineering*. 2025; 30(4):100. https://doi.org/10.1007/s10664-025-10643-z

[82] José Campos; Andriy Miranskyy. Quantum Mutant Equivalence via Transpilation. Preprint. 2026. https://arxiv.org/abs/2606.26604v1 Accessed 19 September 2026.

[83] Pengzhan Zhao, Zhongtao Miao, Shuhan Lan, Jianjun Zhao. Bugs4Q: A benchmark of existing bugs to enable controlled testing and debugging studies for quantum programs. *Journal of Systems and Software*. 2023; 205; 111805. https://doi.org/10.1016/j.jss.2023.111805

[84] Matteo Paltenghi, Michael Pradel. Bugs in Quantum Computing Platforms: An Empirical Study. *Proceedings of the ACM on Programming Languages*. 2022; 6; OOPSLA1; 1-27. https://doi.org/10.1145/3527330

[85] Konrad Wojciechowski, Krzysztof Kurowski, Mateusz Slysz, Tomasz Pecyna. Open QBench: A benchmarking framework for evaluating quantum computing platforms. *Quantum Information Processing*. 2026; 25(3):98. https://doi.org/10.1007/s11128-026-05115-2

[86] Jun Ye; Jun Yong Khoo. EmuPlat: A Framework-Agnostic Platform for Quantum Hardware Emulation with Validated Transpiler-to-Pulse Pipeline. arXiv preprint arXiv:2509.12639v1. 2025. https://arxiv.org/abs/2509.12639v1

[87] Vinooth Kulkarni; Jaehyun Lee; Adam Hutchings; Anas Albahri; Jai Nana; Shuai Xu; Vipin Chaudhary. Efficient Transpilation of OpenQASM 3.0 Dynamic Circuits to CUDA-Q: Performance and Expressiveness Advantages. QCNC 2026; arXiv:2604.11599v1. https://arxiv.org/abs/2604.11599v1

[88] Vlad Stirbu; Arianne Meijer–van de Griend. Observing the Quantum Compiler through Automatic Experiment Tracking for Qiskit. Preprint. 2026. https://arxiv.org/abs/2608.05041v1 Accessed 19 September 2026.

[89] Boshuai Ye; Peng Liang; Maryam Tavassoli Sabzevari; Arif Ali Khan. Auditing Empirical Comparisons in Quantum Software. Preprint. 2026. https://arxiv.org/abs/2607.00516v1 Accessed 19 September 2026.

[90] Qiskit contributors. CommutativeCancellation incorrectly adds spurious global phase to P/U1 gates accumulating to 2π (Issue 16377). *Qiskit issue tracker*. 2026. https://github.com/Qiskit/qiskit/issues/16377 Accessed 13 September 2026.

[91] IBM Quantum. Qiskit 0.45 release notes. *IBM Quantum documentation*. 2023. https://quantum.cloud.ibm.com/docs/en/api/qiskit/release-notes/0.45 Accessed 13 September 2026.

[92] S. Yoo, M. Harman. Regression testing minimization, selection and prioritization: a survey. *Software Testing, Verification and Reliability*. 2012; 22; 2; 67-120. https://doi.org/10.1002/stvr.430

[93] S. Elbaum, A. Malishevsky, G. Rothermel. Incorporating varying test costs and fault severities into test case prioritization. *Proceedings of the 23rd International Conference on Software Engineering. ICSE 2001*. 2001; 329-338. https://doi.org/10.1109/icse.2001.919106

[94] Sebastian Elbaum, Gregg Rothermel, John Penix. Techniques for improving regression testing in continuous integration development environments. *Proceedings of the 22nd ACM SIGSOFT International Symposium on Foundations of Software Engineering*. 2014; 235-245. https://doi.org/10.1145/2635868.2635910

[95] Daniel Elsner, Florian Hauer, Alexander Pretschner, Silke Reimer. Empirically evaluating readily available information for regression test optimization in continuous integration. *Proceedings of the 30th ACM SIGSOFT International Symposium on Software Testing and Analysis*. 2021; 491-504. https://doi.org/10.1145/3460319.3464834

[96] Tomasz Chojnacki, Lech Madeyski. Test case prioritization: A systematic review using snowballing and TCPFramework with approach combinators. *Information and Software Technology*. 2026; 194; 108062. https://doi.org/10.1016/j.infsof.2026.108062

[97] Helge Spieker, Arnaud Gotlieb, Dusica Marijan, Morten Mossige. Reinforcement learning for automatic test case prioritization and selection in continuous integration. *Proceedings of the 26th ACM SIGSOFT International Symposium on Software Testing and Analysis*. 2017; 12-22. https://doi.org/10.1145/3092703.3092709

[98] Jackson A. Prado Lima, Silvia Regina Vergilio. A Multi-Armed Bandit Approach for Test Case Prioritization in Continuous Integration Environments. *IEEE Transactions on Software Engineering*. 2022; 48; 2; 453-465. https://doi.org/10.1109/tse.2020.2992428

[99] Mateusz Machalica, Alex Samylkin, Meredith Porth, Satish Chandra. Predictive Test Selection. *2019 IEEE/ACM International Conference on Software Engineering: Software Engineering in Practice (ICSE-SEIP)*. 2019. https://arxiv.org/abs/1810.05286

[100] Linzhi Huang, Hanyu Pei, Yuechen Li, Beibei Yin, Kai-Yuan Cai. A Dynamic Testing Strategy With Incremental Learning Model for Quantum Programs. *IEEE Transactions on Quantum Engineering*. 2026; 7; 1-27. https://doi.org/10.1109/tqe.2025.3626745

[101] G. Rothermel, R.H. Untch, Chengyun Chu, M.J. Harrold. Prioritizing test cases for regression testing. *IEEE Transactions on Software Engineering*. 2001; 27; 10; 929-948. https://doi.org/10.1109/32.962562